\documentclass[aps,twocolumn,secnumarabic,nobalancelastpage,amsmath,amssymb,nofootinbib]{revtex4}

\usepackage{subfigure}
\usepackage{graphicx}
\usepackage{url}
\usepackage{natbib}

\begin{document}

\title{Annotation Enrichment Analysis: An Alternative Method for Evaluating the Functional Properties of Gene Sets}
\author{Kimberly Glass\,$^{1-3,}$\footnote{contact: kglass@jimmy.harvard.edu}, and Michelle Girvan\,$^{3-5}$}
\affiliation{$^{1}$Department of Biostatistics and Computational Biology, Dana-Farber Cancer Institute, Boston, MA, USA\\
$^{2}$Department of Biostatistics, Harvard School of Public Health, Boston, MA, USA\\
$^{3}$Department of Physics, University of Maryland, College Park, MD, USA\\
$^{4}$Institute for Physical Science and Technology, University of Maryland, College Park, MD, USA \\
$^{5}$Santa Fe Institute, Santa Fe, NM}

\begin{abstract}
Gene annotation databases (compendiums maintained by the scientific community that describe the biological functions performed by individual genes) are commonly used to evaluate the functional properties of experimentally derived gene sets.  Overlap statistics, such as Fisher's Exact test (FET),  are often employed to assess these associations, but don't account for non-uniformity in the number of genes annotated to individual functions or the number of functions associated with individual genes.  We find FET is strongly biased toward over-estimating overlap significance if a gene set has an unusually high number of annotations.  To correct for these biases, we develop Annotation Enrichment Analysis (AEA), which properly accounts for the non-uniformity of annotations.  We show that AEA is able to identify biologically meaningful functional enrichments that are obscured by numerous false-positive enrichment scores in FET, and we therefore suggest it be used to more accurately assess the biological properties of gene sets.
\end{abstract}

\maketitle

\section{Introduction}
\label{Background}
Evaluating the functional properties of gene sets is a routine step in understanding high-throughput biological data \cite{DAVID, MSigDB} and is commonly used both to verify that the genes implicated in a biological experiment are functionally relevant \cite{DAVID} and to discover unexpected shared functions between those genes \cite{Roth03, Morris10}.  One of the most widely used databases for functional annotations is the Gene Ontology (GO) \cite{GO2000, GO2010}.  This database is highly regarded both for its comprehensiveness and its unified approach for annotating genes in different species to the same basic set of underlying functions \cite{GO2000}.  In order to evaluate the strength of connection between a gene signature predicted by an experimental system and the set of genes that are annotated to a given biological function in this database, most functional enrichment analysis tools rely on set-overlap statistics \cite{Rivals07}.  Because these approaches are subject to an increase in type I errors associated with multiple hypothesis testing, corrections such as the Benjamini, Bonferroni, and FDR are often also applied \cite{Khatri05} (discussed in more detail below).

Young et.\ al.\ recently pointed out that these standard statistical approaches are sometimes inadequate, specifically when evaluating the functional properties of gene-sets derived from RNA-seq data since these experiments are prone to selection-bias due to variability in gene length \cite{Young10}.  Despite the wide use of functional analysis tools, however, little attention has been paid to whether or not the underlying properties of the functional databases themselves may contribute to spurious statistical results.  For example, functional terms in GO are related through a directed acyclic graph (DAG) whose structure contributes to the heavy-tailed distribution seen in the number of gene annotations for individual terms \cite{Glass12}.  In this work we investigate whether these annotation properties lead to a bias in the results of traditional functional analysis methods.  We find that the significance level of the association between random gene sets and functional terms in GO are positively correlated with the number of annotations made to the genes in a given gene set.

We also investigate the properties of experimentally-derived gene signatures, as reported in the Gene Signatures Database \cite{GeneSigDB2012} and find that most signatures include a disproportionate number of highly annotated genes.  Furthermore, traditional overlap statistics find significant associations between these signatures and randomly constructed annotation sets.  Consequently, we propose a scheme, called Annotation Enrichment Analysis (AEA), that focuses on the overlap in \emph{annotations} between a set of genes and the set of terms belonging to a branch of the GO hierarchy.  By looking at annotation overlap instead of gene overlap, our approach takes into account the annotation properties of the Gene Ontology.  It effectively eliminates biases due to database construction and highlights relevant biological functions in experimentally-defined gene signatures.  We also provide an analytic approximation to AEA that is able to partially compensate for the biases we find using traditional approaches.  Implementations of both approaches are provided at http://www.networks.umd.edu.

There are many functional annotation databases that have been developed in order to classify genes according their various roles in the cell \cite{COG, KEGG, GenProtEC, MetaCyc, MultiFun}.  Here, we focus our analysis on functional annotations made to the Gene Ontology because of its wide use by many functional enrichment tools (for example \cite{DAVID, MSigDB, GOstat,GOToolBox, topGO}).  Since many of the annotation properties of the Gene Ontology are shared by other databases \cite{Glass12}, we believe that the methods we develop here could be applied to functional enrichment analysis using other classification databases.

The Gene Ontology \cite{GO2000} takes the form of a directed acyclic graph (DAG) in which ``child'' functional categories (``terms'') can be subclassified under one or more other, more general categories, called ``parent'' terms, using ``is a'' and ``part of'' relationships.  ``Branches'' in the Gene Ontology can therefore be defined as sets of terms that contain a parent term and all of its progeny.  Note that these branches will contain overlapping sets of terms since each term can be a descendant of multiple ancestors at each level of the DAG.  Within this structure, genes are annotated to a set of functional categories.  These annotations are transitive such that a parent term will take on all the genes annotations associated with any of its progeny \cite{GO2001}.  Consequently, terms with many progeny often contain many gene annotations whereas terms with few progeny generally have fewer associated genes.  Note that since every parent term takes on the annotations of its progeny, the number of unique genes annotated to a parent term is the same as the number of unique genes annotated to the branch defined by the parent term and its progeny; however, the total number of annotations made to any parent term with children is less than the total number of annotations made to the corresponding branch.  ``Biological Process,'' ``Molecular Function,'' and ``Cellular Component'' are the three most general terms in GO, defining three independent branches such that every other term can only belong to one of these three categories.  As a consequence all genes in GO are annotated to at least one, and often all three, of these categories.

Since we want to determine the influence of annotation properties on functional enrichment analysis, especially in the context of experimental gene signatures in commonly studied diseases such as cancer, we focus our study on GO annotations that are associated with human genes.  With this in mind, we downloaded information regarding gene-term annotations for human genes from the Gene Ontology website (geneontology.org) and used this data to construct a gene-term bipartite graph, represented as an $n_G \times n_T$ adjacency matrix, where $n_G$ is the total number of genes and $n_T$ is the total number of terms listed in the annotation file.  In this matrix a value of one indicates a known connection between the corresponding gene and term, and a value of zero indicates that the gene is not associated with that term.

\begin{figure}[!tpb]
\subfigure[Term Degree Distribution]{\includegraphics[width=120px]{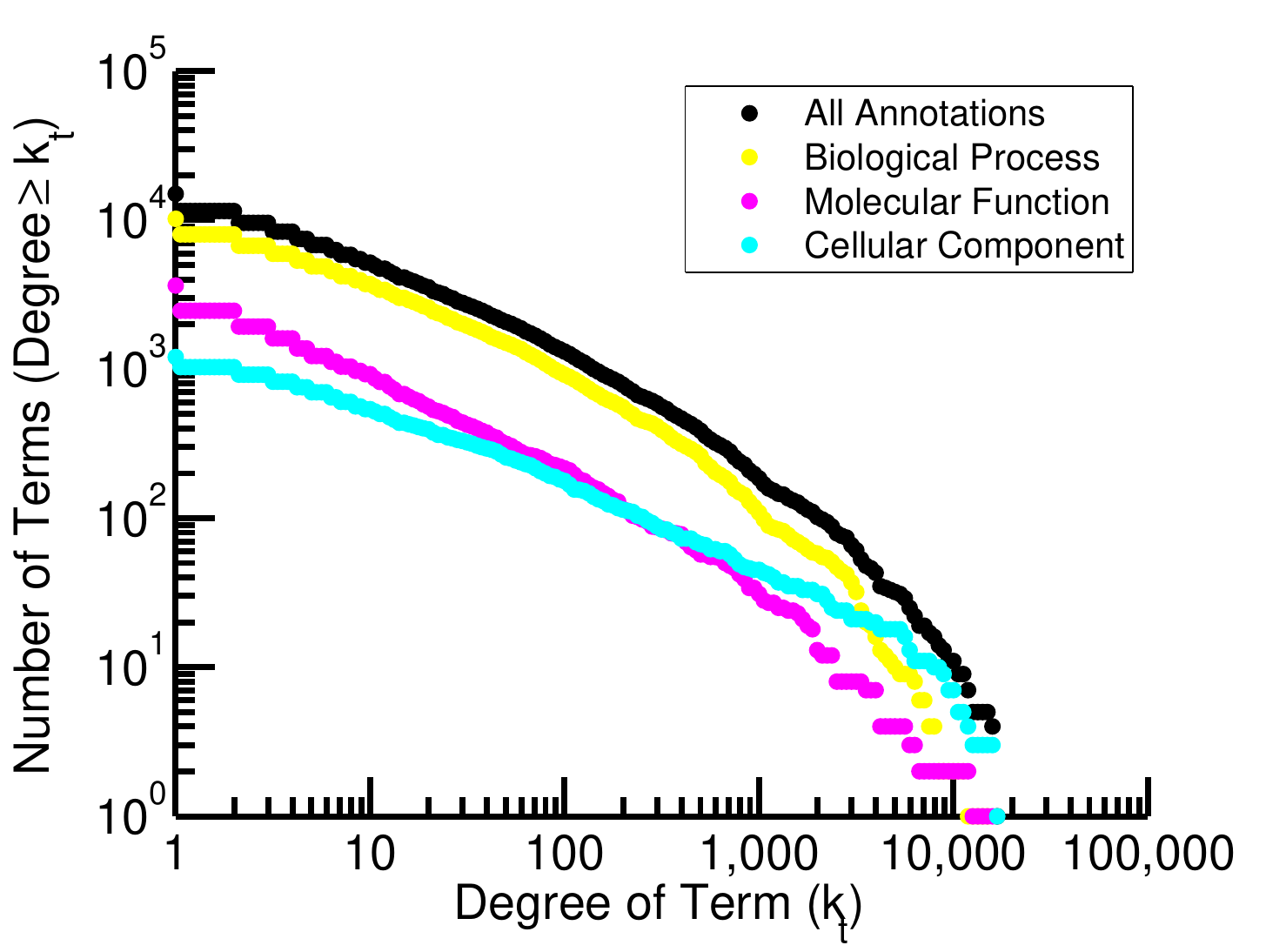}\label{TermDegreeDist}}
\subfigure[Gene Degree Distribution]{\includegraphics[width=120px]{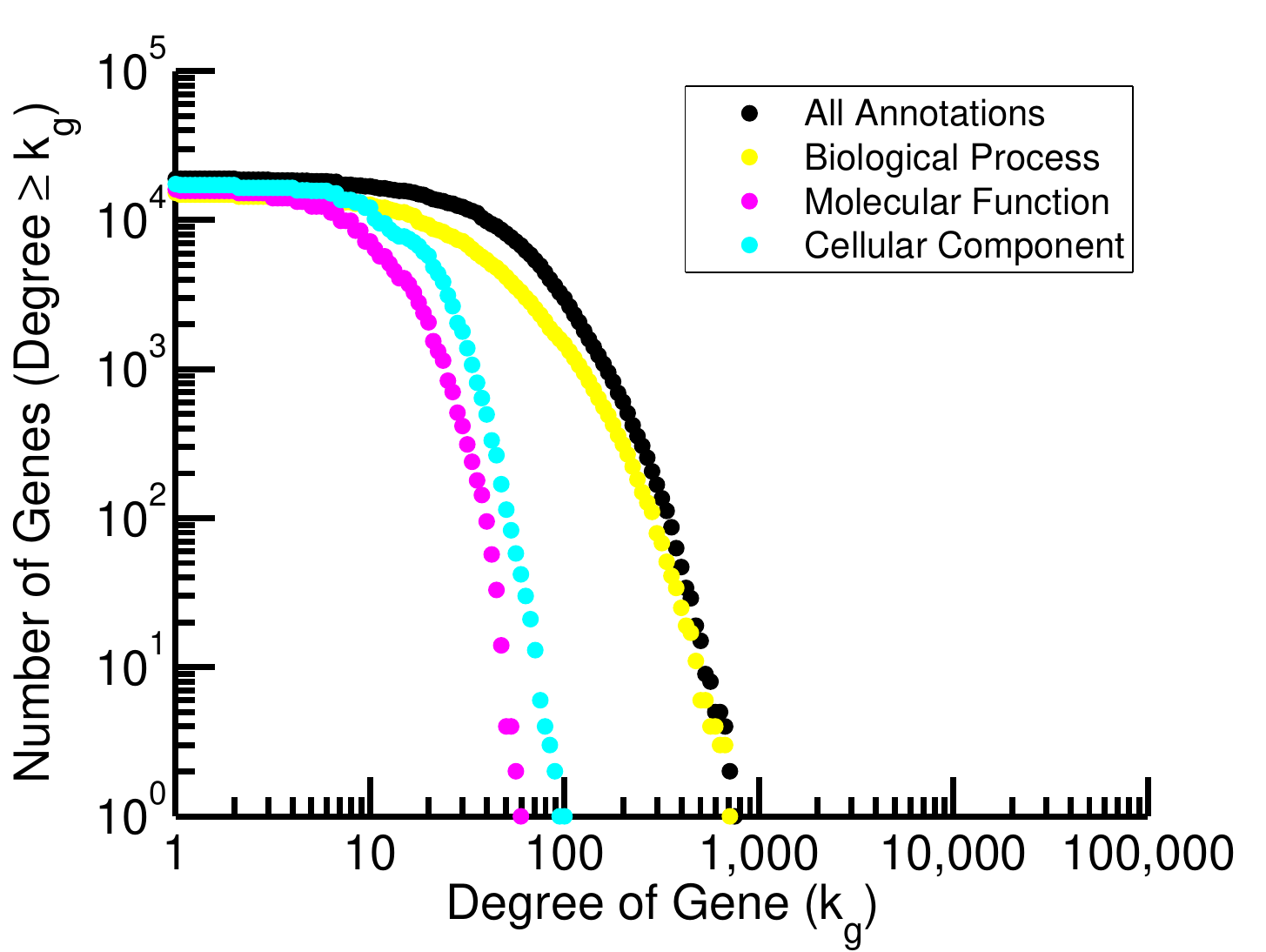}\label{GeneDegreeDist}}
\caption[Annotation Properties of the Gene Ontology]{The cumulative degree distributions of (a) genes and (b) terms in human GO annotations.}
\label{Figure1}
\end{figure}

In this bipartite graph many terms are only associated with a small handful of genes, while some terms are associated with many genes.  A histogram of the ``degree'' of terms ($k_t^{(p)}$, or the number of genes annotated to term $p$) reveals a heavy-tailed relationship (Figure \ref{TermDegreeDist}).  In contrast, a histogram of the ``degree'' of genes ($k_g^{(i)}$, or the number of terms to which gene $i$ is annotated) shows that although some genes have many more annotations than others, the distribution is not as skewed as the term degree distribution. (Figure \ref{GeneDegreeDist}).

The ``Biological Process'' ontology contains a significant fraction of the total annotations.  Although all three ontologies are used in functional enrichment analysis, it is common to focus on this ontology, both for its size and because its members describe dynamical processes performed by the cell.  We will do the same in the following analysis.  The total number of annotations made to the ``Biological Process'' ontology is $656783$, originating from $18930$ genes to $10192$ terms.  Consequently, the average number of annotations made by an individual gene is $43.2$ and the average number of annotations made to an individual term is $64.4$. These values will be useful to keep in mind, especially as we investigate the annotation properties of gene signatures and of the terms for which they are enriched.

The most widely used statistics for evaluating which functional categories are enriched in a set of genes are based on gene counts and include Fisher's Exact Test, the binomial test, and the chi-squared test \cite{Rivals07}.  Although these statistics vary in exact implementation, they all rely on the same basic underlying assumption that all genes have an equal probability of being selected under the null hypothesis.  Of these tests, Fisher's Exact Test (FET) is the most common statistic and is used by many of the most popular functional enrichment tools (see Table 2 in \cite{Khatri05}), and therefore we choose it to represent a ``typical'' evaluation of gene set functional enrichment.  Although it is recognized that this statistic makes assumptions in its null hypothesis that fail to reflect the complex properties of the Gene Ontology, it is widely regarded as a good guide in determining what types of functions are represented in a given set of genes.

Since most functional enrichment analysis compares a gene set to all the terms in GO, multiple-hypothesis testing corrections are often applied to these p-values \cite{Khatri05}.  These corrections raise the value at which a comparison between a gene set and a GO term should be considered significant.  Commonly used multiple-hypothesis corrections include the Bonferroni, Benjamini and the False Discovery Rate.  Of these, the Bonferroni is the most conservative and adjusts the value at which a test is considered ``significant'' by the number of tests made.  However, although this correction will change the critical value of individual tests, but will not affect the rank ordering of these tests.  In contrast, the False Discovery Rate (FDR) adjusts the value at which a test is considered ``significant'' based on the rank of the predicted level of significance.  It will provides approximately the same correction as the Bonferroni for the most significantly-ranked p-values but will not adjust tests that are the least-significant by rank.  As a consequence, the rank ordering of the significance can change slightly when using the FDR.  For more details on how to calculate these scores, see Methods.

\section{Results}

\subsection{Annotation Properties Influence the Results of Functional Enrichment Analysis}
\label{EffectOfAnnotationBias}

We wished to determine the effect of annotation properties on functional enrichment analysis.  First, we created we created random gene sets with $N_g=200$ members each, but in which we controlled the total number of annotations ($M_g$) made by the genes belonging to each gene set, such that the average degree of the genes in the set ($k_{avg}=M_g/N_g$) varies from approximately $21$ to $65$, or from around half to $1.5$ times the expected average degree of $43$.  Next, we also created random ``branches''.  We note that because terms in GO are related via a hierarchical DAG, they can often share many of the same gene annotations.  Therefore, our random ``branches'' are composed of random selections of GO terms whose gene annotations will collapse together such that their ``faux'' parent term will have the same total number of gene annotations as a term that currently exists in GO.  In other words, each random ``branch'' will have the same \emph{number} of unique genes annotated to it as a real GO branch, but these gene annotations to the faux parent term will be influenced by a cumulation of annotations made to a random set of progeny terms.  For details on how we constructed these random gene sets and ``branches'' see Methods.

\begin{figure}[!tpb]
\subfigure[Fisher's Exact Test (GO Branches)]{\label{FETVaryK}\includegraphics[height=90px]{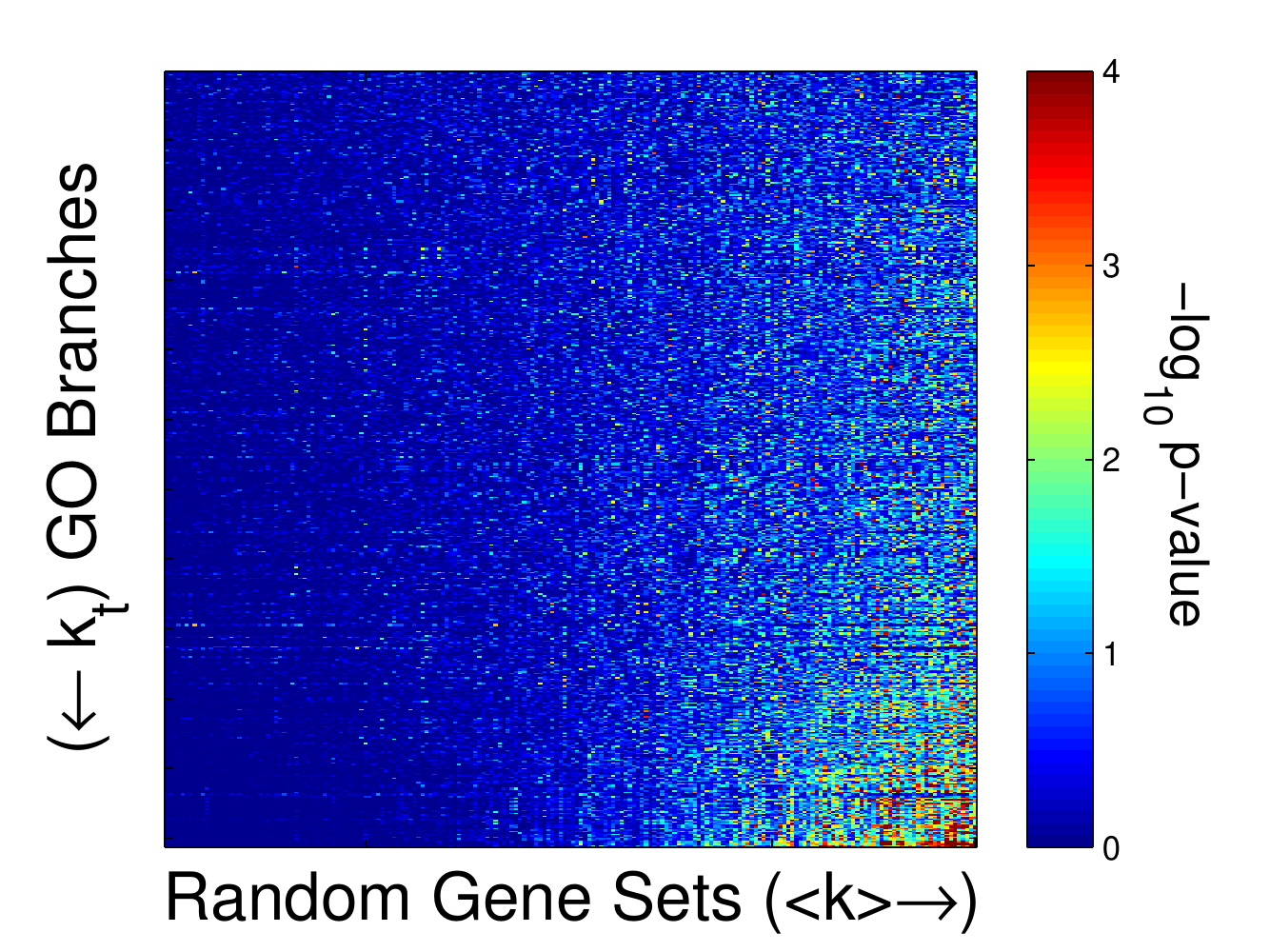}}
\subfigure[Fisher's Exact Test (random ``Branches'')]{\label{FETVaryKbranch}\includegraphics[height=90px]{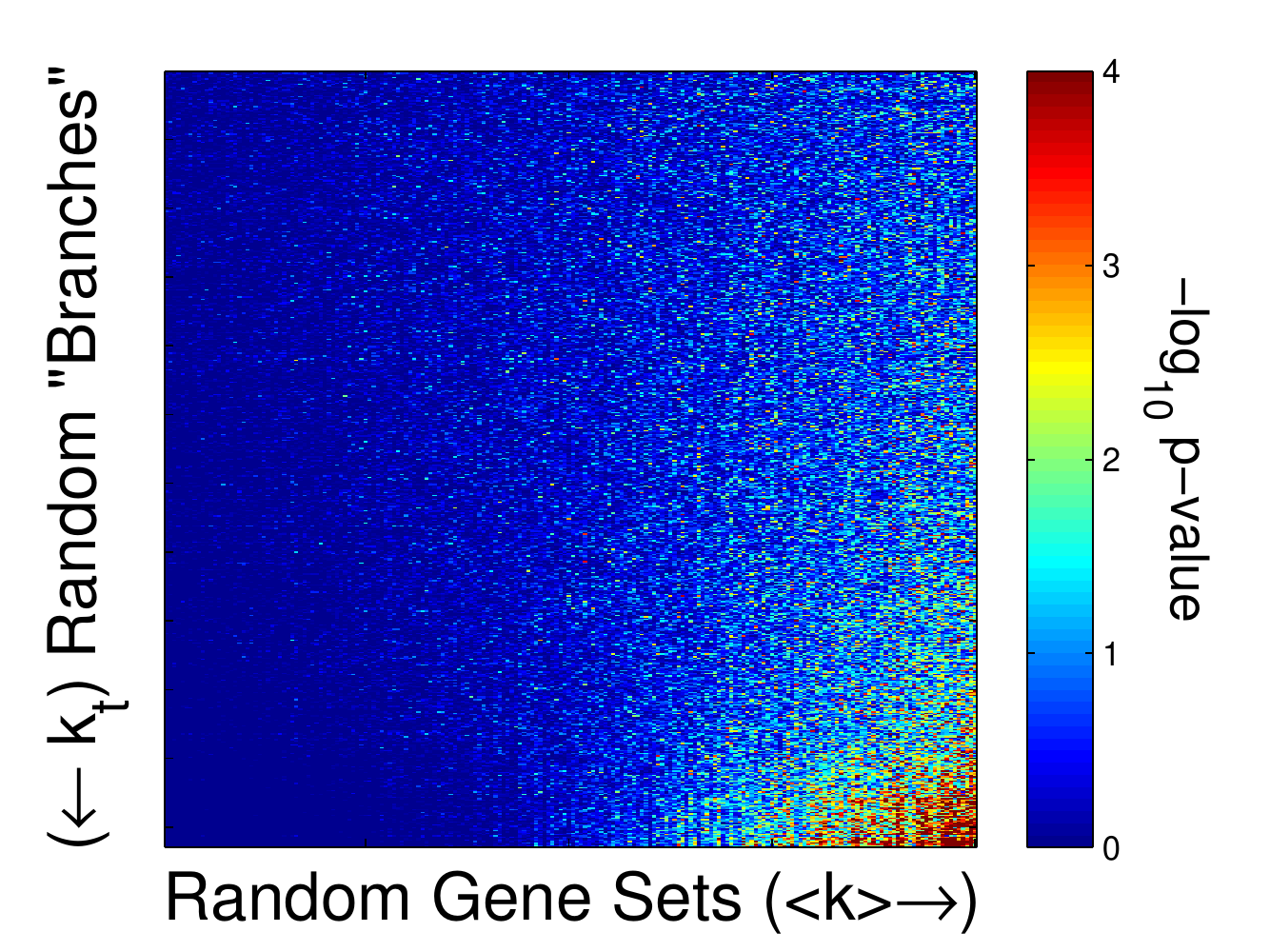}}
\subfigure[Annotation Enrichment Analysis (GO Branches)]{\label{AEAVaryK}\includegraphics[height=90px]{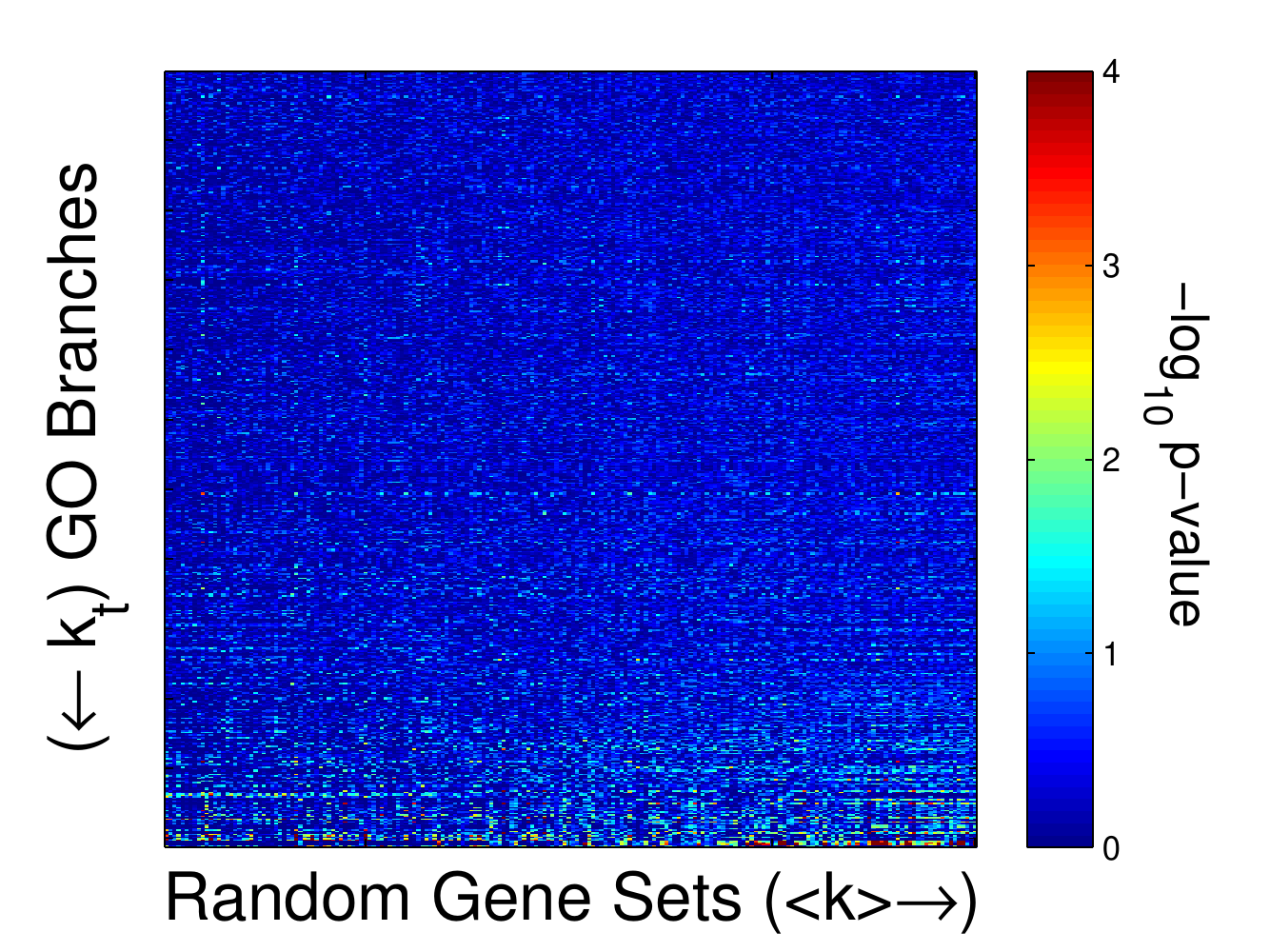}}
\subfigure[Annotation Enrichment Analysis (random ``Branches'')]{\label{AEAVaryKbranch}\includegraphics[height=90px]{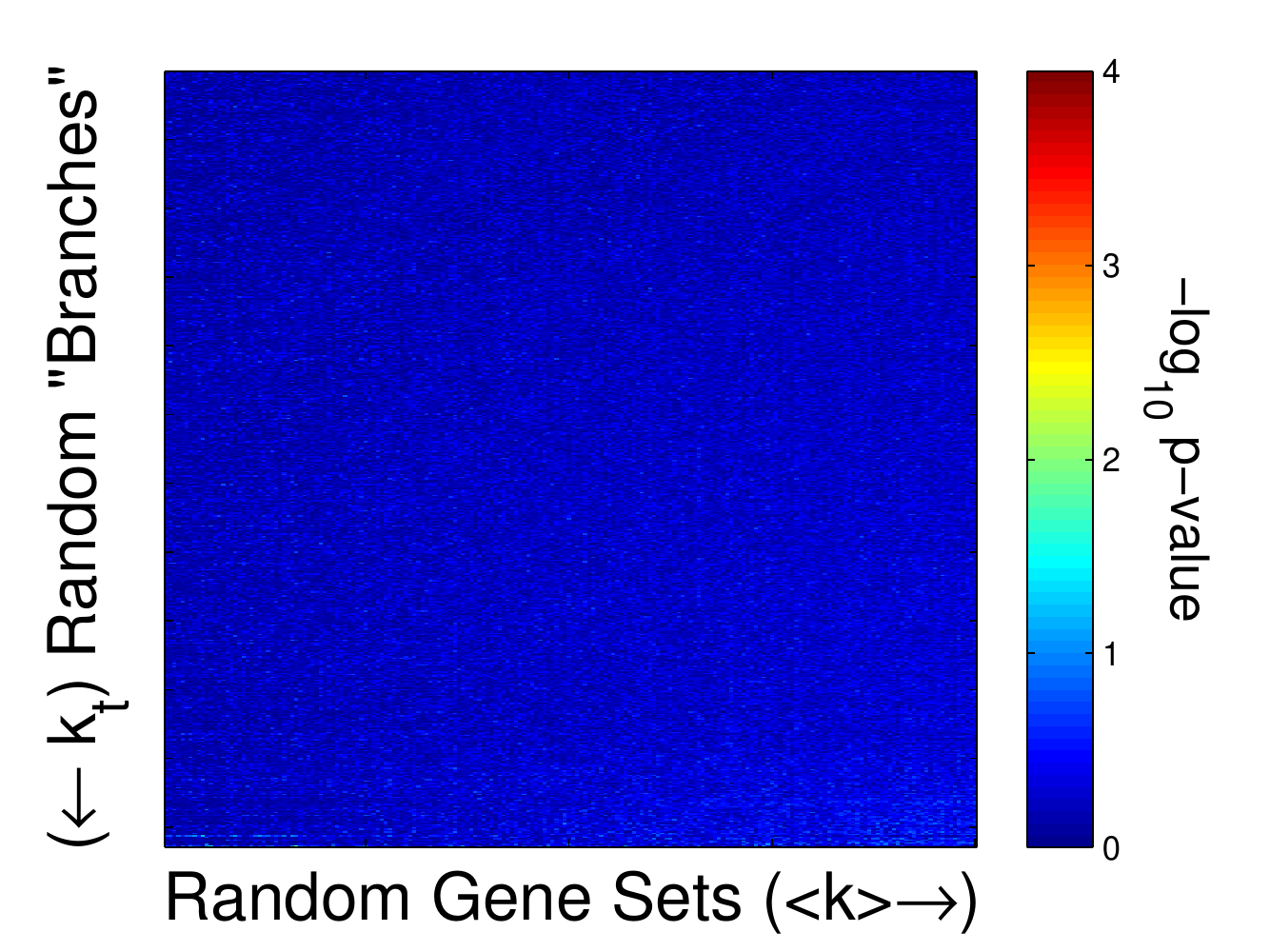}}
\caption[Evaluating and Correcting for the Effect of Annotation Properties In Functional Analysis]{The enrichment (measured by p-value) of 200 randomly generated gene sets in annotations made to either real GO branches or randomly constructed ``branches''.  The branches are ordered based on how many genes are annotated to the parent term ($k_t$) and the gene sets are ordered based on total the number of annotations ($M_g$) made by the $200$ genes in that set.  There is an obvious bias toward significant enrichment between high degree gene-set/term pairs in (a) Fisher's Exact Test (FET), (b) even when using randomly constructed ``branches''; however this is correctly accounted for using (c-d) Annotation Enrichment Analysis (AEA).}
\label{Figure2}
\end{figure}

As an initial test, we used our random gene sets to evaluate how annotation bias might effect the enrichment significance predicted in a functional enrichment test.  To do this we used FET to determine the enrichment of our randomly constructed gene sets in GO terms from the ``Biological Process'' ontology.  Figure \ref{FETVaryK} shows the results for terms that have $200$ or more gene annotations, ordered based on their total number of gene annotations.  The trend is striking.  Even though they have the same number of members, gene sets with a higher number of \emph{annotations} are more enriched in GO terms compared to gene sets with a lower number of annotations.  Branches with an increased number of genes annotated to the parent term ($k_t$) also tend to be the most significantly enriched, especially in these ``highly-annotated'' gene sets, consistent with that fact that the significance level predicted by a method such as FET is dependent on the number of members in a given gene set.  We point out that although multiple-hypothesis corrections will sufficiently raise a p-value such that either very few or no false positives will occur, the biases themselves cannot be overcome in this manner.  A Bonferroni correction will not chance the ordering of the p-values and the bias will remain.  Even an FDR correction, which can alter the rank ordering of significance, is insufficient to overcome this strong signal (see Supplemental Figure \ref{SFig1}).

Next, we investigated what effect, if any, overlapping annotations between GO terms might have on functional enrichment analysis.  Namely, we used FET to determine the enrichment of our random constructed gene sets in our random ``branches''.  Interestingly, the trend we observed with real GO branches -- that those with more gene annotations to the parent term also tend to be enriched in ``high-degree'' gene sets -- appears to be even more pronounced when using random ``branches'' (Figure \ref{FETVaryKbranch}).  This suggests that part of the bias from FET might result from overlapping annotations to GO terms.

\subsection{Annotation Enrichment Analysis Corrects for Annotation Bias}
\label{AEAMethod}
\begin{figure}[!tpb]
\includegraphics[width=240px]{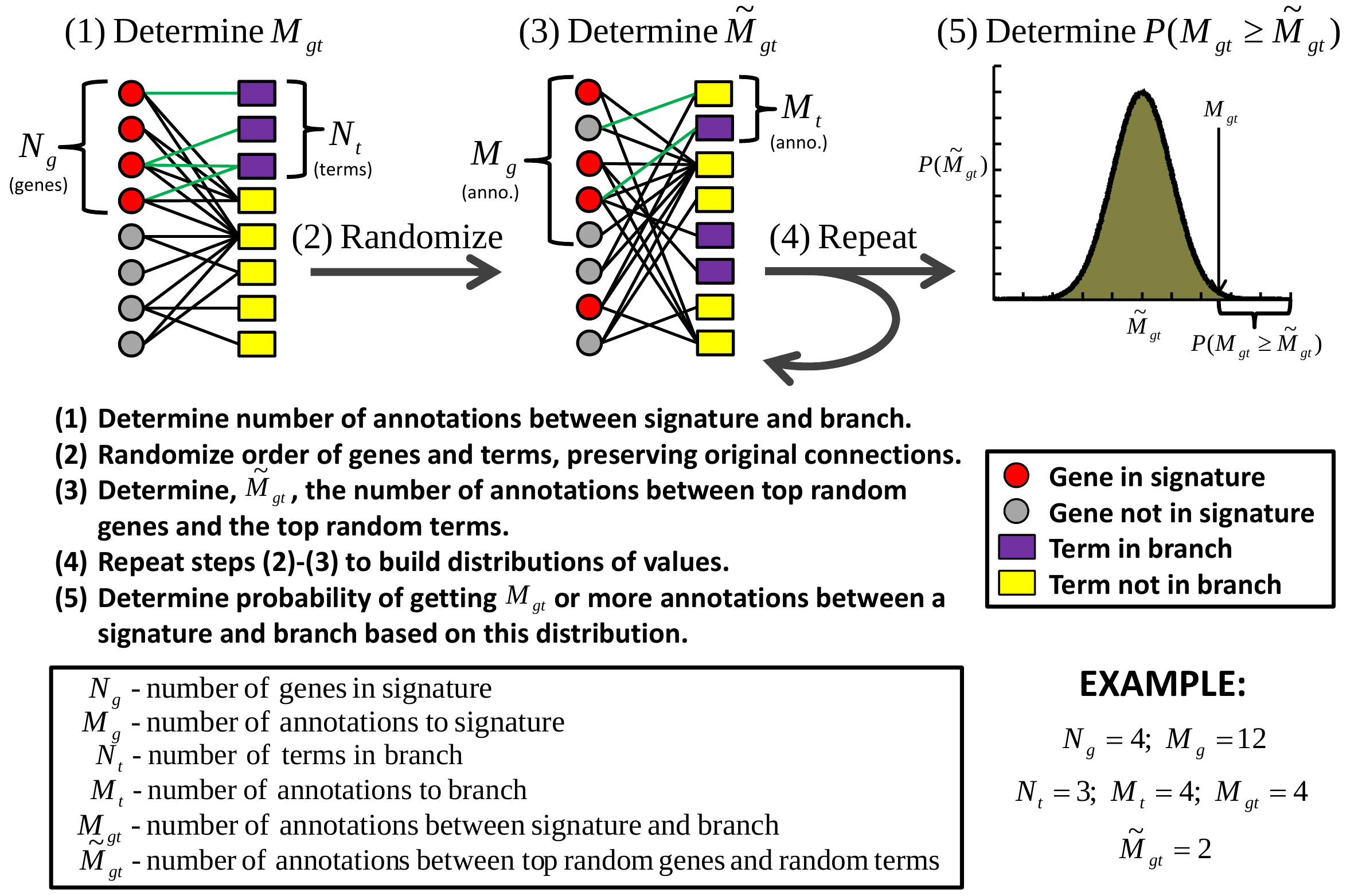}
\caption[Annotation Enrichment Analysis]{An outline of how Annotation Enrichment Analysis (AEA) calculates the significance of association between a given gene signature and the collection of terms that belong to a branch in the GO hierarchy.}
\label{Figure3}
\end{figure}

Clearly annotation properties of both genes and functional categories can influence the results of functional enrichment analysis.  In order to mitigate these effects, we suggest that instead of considering the overlap between two gene sets, as is traditionally done in functional enrichment analysis, one instead considers the overlap between \emph{annotations} made to a gene set and a branch of terms in the Gene Ontology.  To accurately capture the significance of annotation overlap we develop a randomization scheme that preserves the structure of GO annotations while calculating the probability of obtaining a certain number of co-annotations between a gene set and a GO branch.  We call this approach Annotation Enrichment Analysis (AEA) and illustrate it in Figure \ref{Figure3}.

In this randomization is it useful to think of the Gene Ontology as a bipartite graph (see Introduction).  We begin by determining $M_g$, the number of annotations to a gene set, $M_t$, the number of annotations to the terms in a GO branch, and $M_{gt}$, the number of annotations stretching between this gene set and branch.  We then determine a distribution for the expected number of co-annotations.  To do this we, simultaneous, randomly permute the order of genes and terms while still preserving the original connections from the GO bipartite graph.  We then take annotations connected to the top random genes until we've selected $M_g$ annotations, and annotations connected to the top random terms until we've selected $M_t$ annotations, and determine $\tilde{M}_{gt}$, the number of edges in the bipartite graph that extend between these top random genes and top random terms.  In the (fairly common) case where selecting the top $M_g$/$M_t$ annotations does not correspond to selecting a whole number of genes/terms, we take the top number of genes/terms whose total annotations is closest to $M_g$/$M_t$, respectively.  We repeat the randomization process many times in order to determine a distribution of values for $\tilde{M}_{gt}$.  We define a new p-value, $p_A(M_{gt})$ which reflects the probability that $\tilde{M}_{gt} \geq M_{gt}$:
\begin{equation}
p_{A}(M_{gt}) = P(\tilde{M}_{gt} \geq M_{gt}).
\label{AEAequation}
\end{equation}

We determined the significance of all GO branches in our randomly generated gene sets with AEA (using $10^4$ randomizations), and created a heat map of these values as we had done for the significance values produced using standard set-overlap statistics (Figure \ref{AEAVaryK}). The results of AEA are uniform across varying gene set degree, demonstrating that AEA works well at eliminating annotation bias.  We also determined the significance of all random ``branches'' in our randomly generated gene sets with AEA ($10^4$ randomizations), and created a heat map of these values (Figure \ref{AEAVaryKbranch}).  Whereas the FET results looked nearly identical, the significance predicted by AEA is much lower for the randomly generated ``branches'' compared to the real GO branches.

\subsection{Experimental Gene Signatures are Often Highly-Annotated}
\label{SigProperties}
One of the most common applications of enrichment analysis is to ascertain the functional properties of an experimentally determined set of genes.  Although we have demonstrated that AEA corrects for annotation bias with randomly generated gene sets, we also want to know how well this analysis can recapitulate biologically-relevant results.  With this in mind we downloaded signatures as recorded in the Gene Signatures Database (GeneSigDB) \cite{GeneSigDB2012}.  This database is a manual curation of previously published gene expression signatures, focusing primarily on cancer and stem cell signatures \cite{GeneSigDB2010}.  In the following analysis we will use all $309$ human signatures from this database that contain at least $100$ and less than $1000$ genes that also are annotated to a term in the ``Biological Process'' ontology.

\begin{figure}[!tpb]
\subfigure[Signature Properties]{\label{SizeVsK}\includegraphics[height=90px]{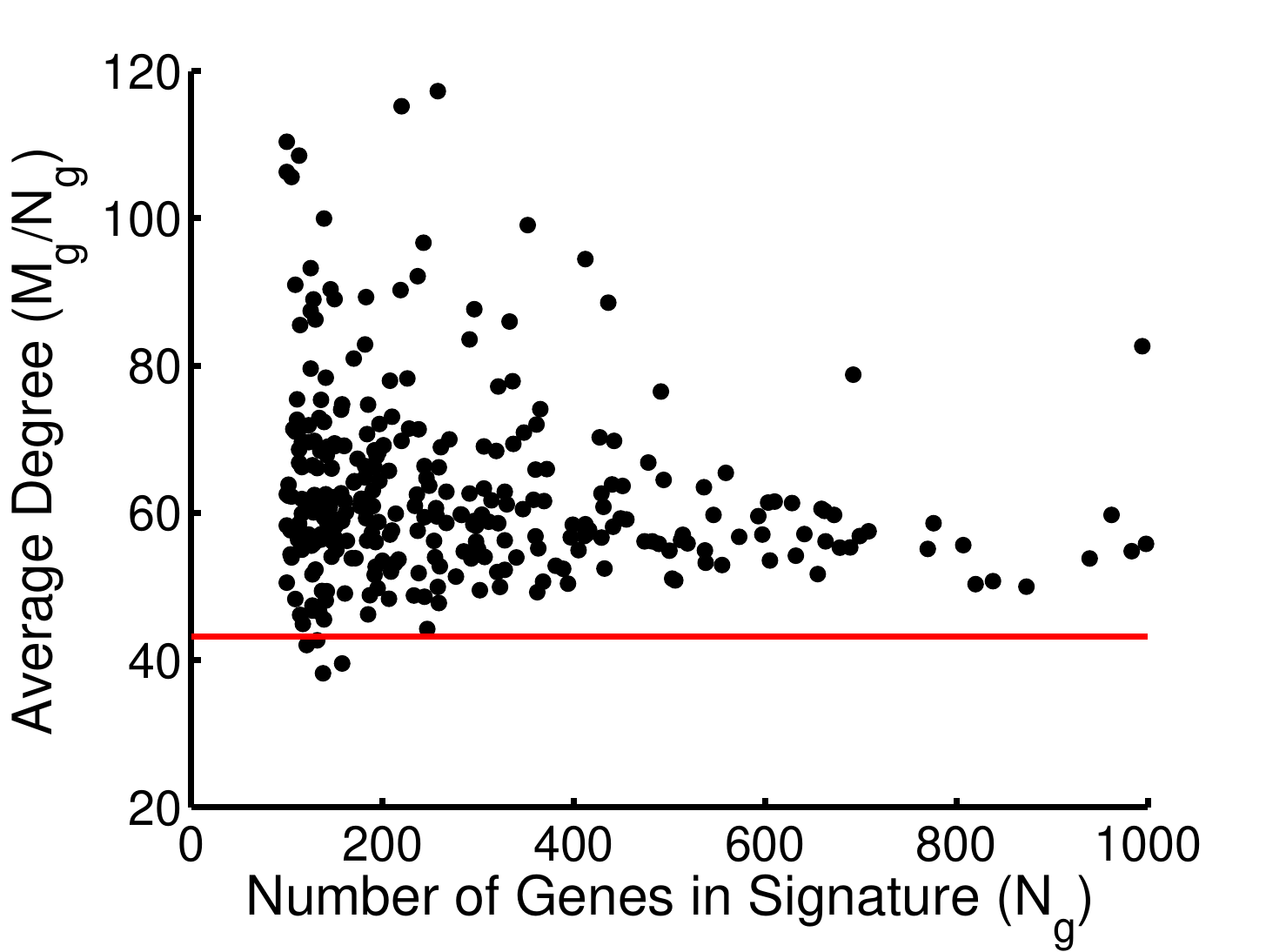}}
\subfigure[FET vs. AEA on Signatures]{\label{PValCompare}\includegraphics[height=90px]{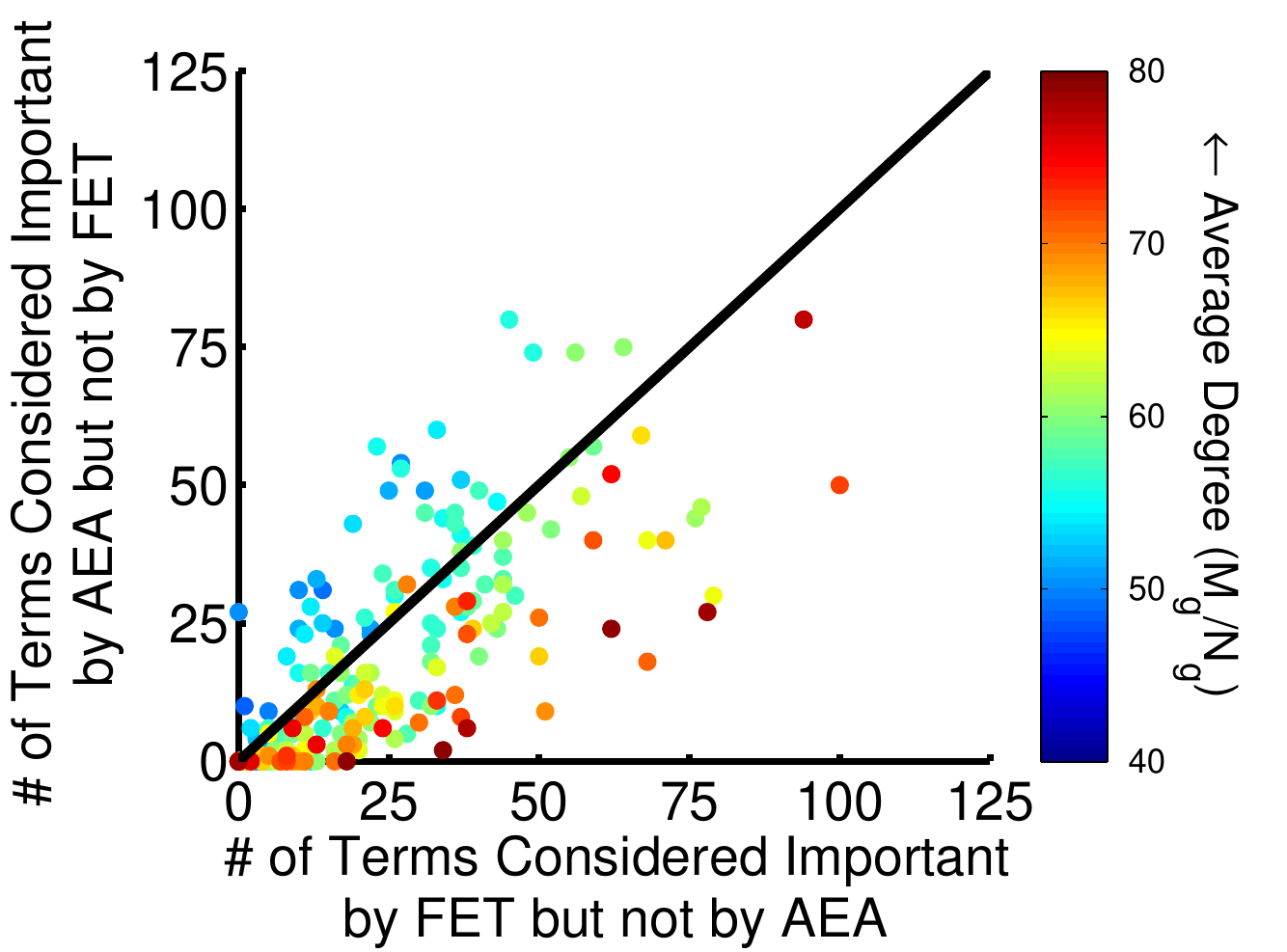}}
\caption[Annotation Properties of Experimental Gene Signatures]{Annotation properties of experimental gene signatures.  (a) The number of genes versus the average number of annotations made to the genes in each signature.  Genes from signatures generally contain many more GO annotations than one would expect if selecting genes randomly (red line).  (b) The number of terms that are considered important (top 10\% by rank) by one of the measures (either AEA or FET), but not important (bottom 80\% by rank) by the other, plotted for each gene signature.  The signatures are colored according to the average level of annotation ($k_{avg}=M_g/N_g$).}
\label{Figure4}
\end{figure}

First, to assess whether annotation bias might play a role in evaluating the functional properties of these gene signatures, we determined the average number of annotations made to the genes occurring in each signature. Figure \ref{SizeVsK} shows the number of genes in a signature plotted against the average level of annotation for each signature.  The expectation for a random selection of genes (the average number of annotations made to all genes -- see Introduction) is shown as a red line.  The plot suggests that many genes belonging to these signatures are also more highly annotated in GO.  Almost a third ($99$) of the signatures have an average level of annotation that is greater than any of our randomly generated gene sets and all but four signatures have an average level of annotation greater than expected by chance.  Since we have shown that random gene signatures with these annotation levels encounter a bias in traditional functional enrichment analysis, we believe these experimental signatures are an appropriate biological set with which to evaluate how AEA compares to FET when investigating and discovering the functions of genes contained in experimental biological data.

Therefore, next we predicted the enrichment of all ``Biological Process'' GO terms in these signatures both by traditional set-overlap statistics (FET) as well as with AEA ($10^{6}$ randomizations).  We first tested to see if the two measures gave the same general results; in other words, are the categories ranked highly by FET also ranked highly by AEA and are the categories ranked poorly by FET also ranked poorly by AEA.  To this end we selected the top $10\%$ of terms based on their enrichment score in FET and AEA to designate as ``important'' according each to these measures.  We compared this list of terms to the list of terms that are ``not important'' (in the bottom $80\%$ of terms by rank) according to each measure.  The number of terms considered important in AEA but not by FET versus the number of terms considered important by FET but not AEA for each signature is plotted in Figure \ref{PValCompare}.  Complete agreement between FET and AEA on this plot is represented by a point at $(0,0)$, and complete disagreement is represented by a point at $(1019,1019)$. In order to see how annotation properties influenced any differences, we colored signatures based on the average level of annotation to their member genes.

There is some agreement between AEA and FET, as many points fall fairly close to the origin and, at most, reflect only a $10\%$ difference in identified ``important'' terms.  However, annotation bias is evident.  In signatures containing the highest levels of annotation, the terms deemed most ``important'' by FET are more likely to be considered ``unimportant'' according to AEA, and vice versus.  These results are consistent with the previous analysis in random gene sets that showed a bias by FET to place more significance between gene sets and terms with a higher number of annotations (see Figure \ref{Figure2}).  It also demonstrates that annotation bias is present when evaluating experimentally-derived gene signatures and it not an artifact of how we constructed our random gene sets.

\subsection{Annotation Enrichment Analysis Uncovers Meaningful Biological Associations}
\label{AEABiology}

Next, we investigated the specific biology that is highlighted using AEA and FET.  We chose approximately forty representative terms/signatures for each measure by ranking according to the minimum enrichment score each has across all signatures/terms and selecting the most ``significant'' terms/signatures by this rank.  For AEA a few ($981$ out of $3149328$ possible) term-signature pairs have an estimated p-value of $p<10^{-6}$ after one million randomizations, therefore, when necessary, we broke ties by the number of signatures/terms enriched in the terms/signatures at this level.  We performed hierarchical clustering (using the ``clustergram'' function in Matlab with default settings) on the terms and signatures selected for each measure.  The results are shown in Figure 5.

\begin{figure*}[!tpb]
\subfigure[Fisher's Exact Test]{\label{FETcluster}\includegraphics[height=250px]{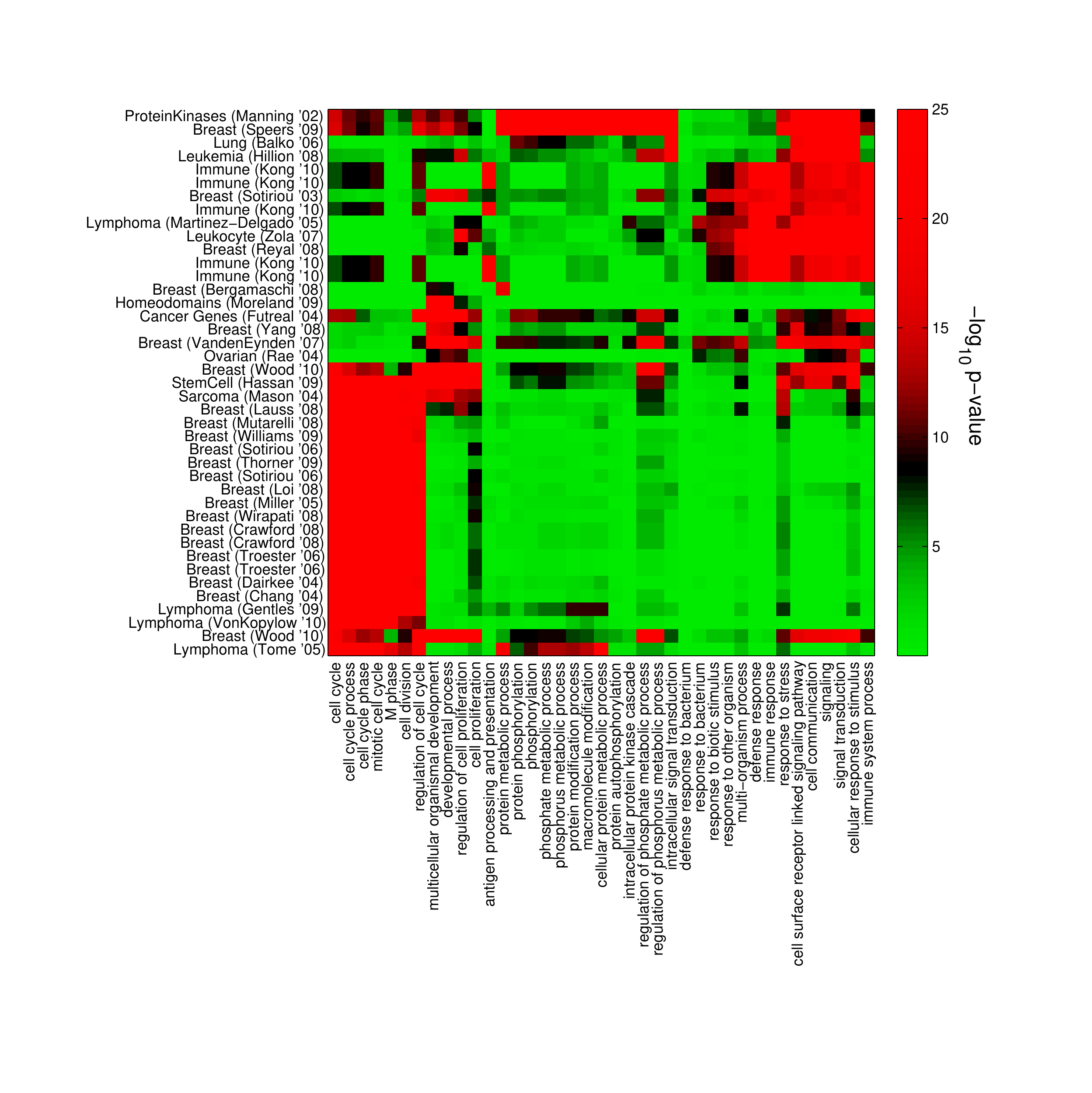}}
\subfigure[Annotation Enrichment Analysis]{\label{AEAcluster}\includegraphics[height=250px]{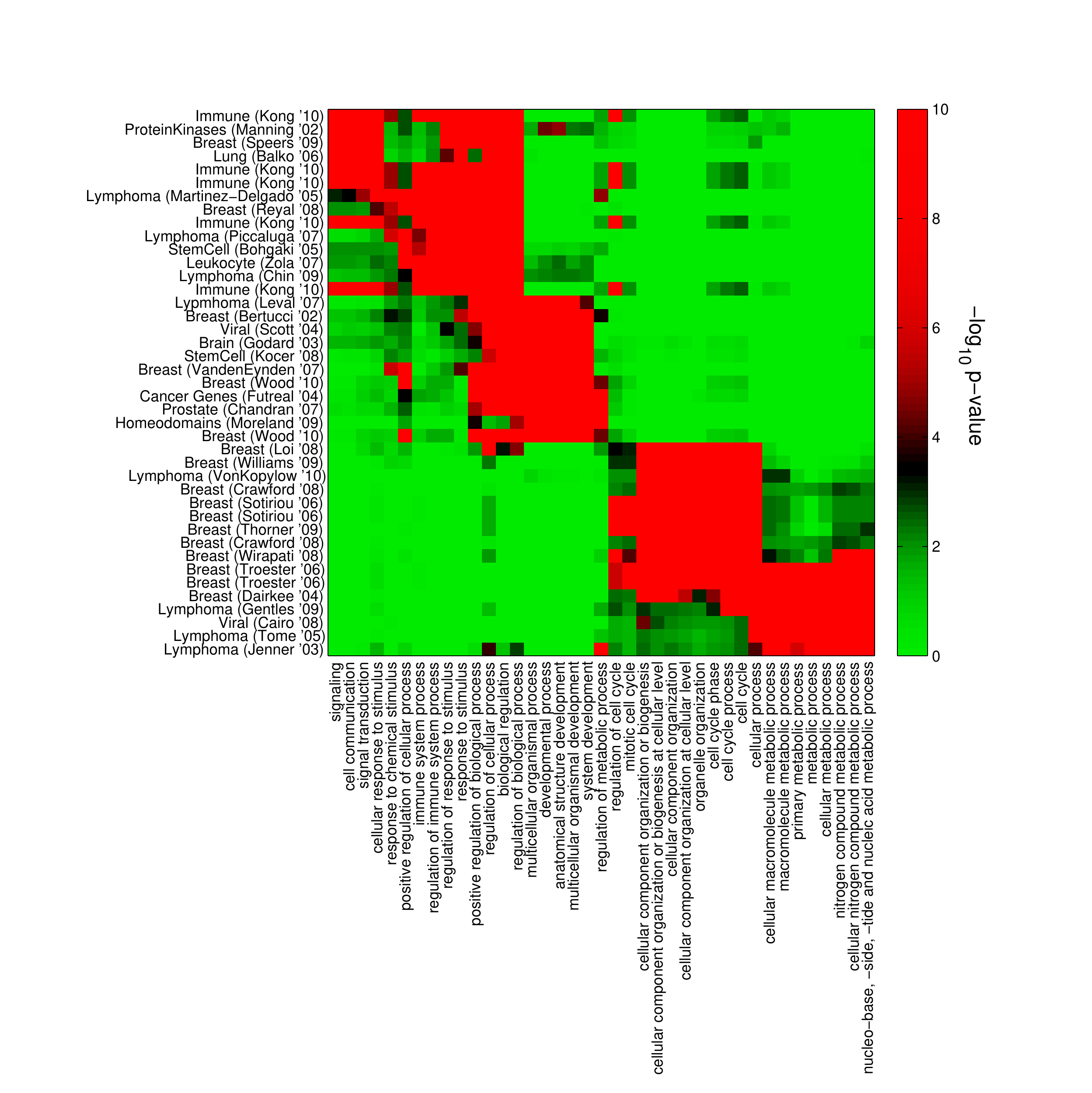}}
\caption[Summary of Functional Results Predicted by FET and AEA in Experimental Gene Signatures]{Clustergrams representing enriched term-signature pairs. (a) A clustering of signatures and terms selected based on their enrichment-score according to FET.  These signatures include those reported in \cite{Manning02, Speers09, Balko06, Hillion08, Kong10, Sotiriou03, MartinezDelgado05, Zola07, Reyal08, Bergamaschi08, Moreland09, Futreal04, Yang08, VandenEynden07, Rae04, Wood10, Hassan09, Mason04, Lauss08, Mutarelli08, Williams09, Sotiriou06, Thorner09, Loi08, Miller05, Wirapati08, Crawford08, Troester06, Dairkee04, Chang04, Gentles09, VonKopylow10, Tome05}.  (b) A clustering of signatures and terms selected based on their enrichment-score according to AEA.  The signatures and terms break into several, biologically distinct units.  One is associated with immune-response, and includes signatures published in \cite{Kong10, Manning02, Speers09, Balko06, MartinezDelgado05, Reyal08, Piccaluga07, Bohgaki05, Zola07, Chin09}.   A second includes signatures related to cellular-differentiation published in \cite{Leval07, Bertucci02, Scott04, Godard03, Kocer08, VandenEynden07, Wood10, Futreal04, Chandran07, Moreland09}. Another cluster includes breast cancer signatures published in \cite{Loi08, Williams09, VonKopylow10, Crawford08, Sotiriou06, Thorner09, Wirapati08, Troester06, Dairkee04}.  Finally three lymphoma \cite{Gentles09, Tome05, Jenner03} and a viral signature \cite{Cairo08} associated with proliferation are also included.  The colorscales for the p-values were chosen to give approximately the same red/green balance in each clustergram.}
\label{Figure5}
\end{figure*}

Clustering the FET results gives rise to a weak visual segregation of terms and signatures into groups (Figure \ref{FETcluster}).  These groups highlight the relationship between the gene signatures and several important biological processes.  For example, the FET clustering shows an enrichment of cell-cycle related processes in breast cancer signatures \cite{Loddo09} and includes immune-related terms enriched in immune gene signatures.  These two groups, however, account for only about half of the selected terms; the clustergram also includes a number of functional categories related to ``proteins'' and ``phosphorylation'' that are only enriched in a small number of signatures.  From this analysis we suggest that the results of FET might be muddled by a signal driven by annotation bias, highlighting either highly-annotated signatures or more general biological processes.

In contrast, using AEA, distinct clusters of signatures and terms emerge (Figure \ref{AEAcluster}). The first includes signatures from immune-systems, lymphoma and leucocytes, and is logically also enriched in terms such as ``immune system'' and ``response to stimulus'' as well as terms related to ``biological regulation''.  Interestingly, one of the breast signatures associated with this cluster \cite{Reyal08} represents a list of genes defined based on immune response in breast cancer and the stem cell signature \cite{Bohgaki05} is from a study on patients with systemic sclerosis, a type of autoimmune disorder.  In addition, the inclusion of a protein-kinase signature \cite{Manning02} is interesting as MAP kinases have been shown to play an important role in immune response \cite{Dong02}.

Another cluster is enriched in categories such as ``system development'' and ``developmental process'' and includes several signatures associated with stem cells or identified based on their role in cellular differentiation.  It also includes a signature of oncogenes \cite{Futreal04}, as well as a signature of homeodomain proteins, known to initiate cascades of genes that in turn will induce cellular differentiation into tissues and organs (e.g. \cite{citeulike:12015164,citeulike:12015165}).  The next cluster, associated primarily with breast cancer signatures, shows a strong enrichment for terms related to the cell cycle and cellular component organization, processes known to be differentially regulated in breast cancer \cite{Loddo09}.  Finally, two lymphoma and one viral signature that were identified based on cell proliferation (for example, by association with Myc targeting \cite{Gentles09, Cairo08}) are enriched for terms such as ``cellular metabolic process.''  This is consistent with expectation since there is evidence that a connection exists between proliferation and metabolic pathways in cancer cells \cite{DeBerardinis08,VanderHeiden09}.

\subsection{Some Predictions Made by FET are Likely a Consequence of Annotation Bias in Experimental Gene Signatures}

\begin{figure*}[!tpb]
\includegraphics[width=500px]{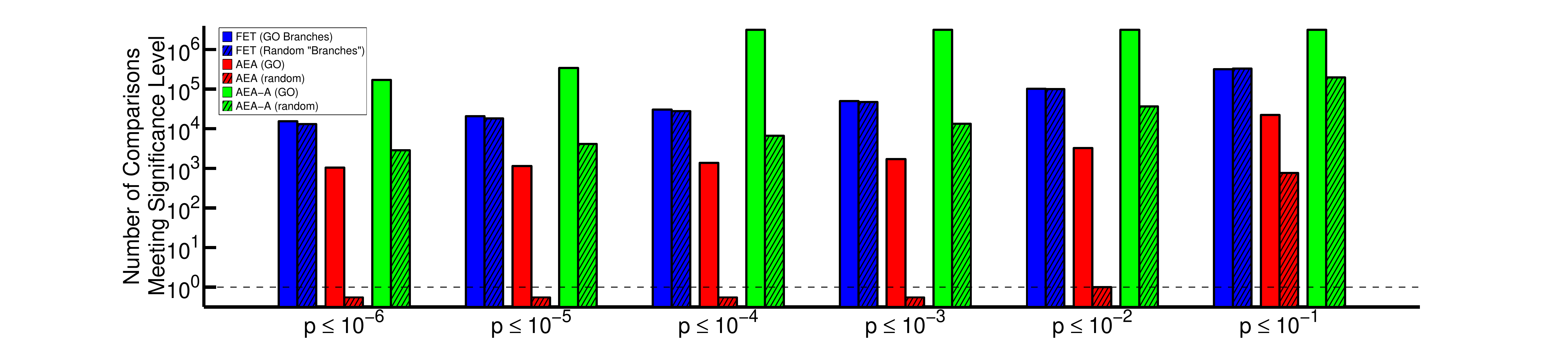}
\caption[Effects of Overlapping Term Annotations]{A plot of the number of term-signature comparisons deemed ``significant'' at various p-value thresholds.  A dotted line indicates only one comparison with a p-value less than or equal to the indicated threshold, and cases where no significant comparisons were found for the corresponding p-value are indicated by a bar not exceeding this line.  Evaluations using gene annotations to GO branches are shown as solid colors, whereas evaluations using genes annotated to random ``branches'' are striped.}
\label{Figure6}
\end{figure*}

Finally, we also wanted to determine how database construction might effect the results of functional enrichment analyses using experimental gene signatures.  Therefore, we investigated the enrichment of experimental gene signatures in our randomly constructed ``branches'' and compared the results with that from real GO branches.  Briefly, for each evaluation, we determined the number of term-signature comparisons considered significant at several different thresholds and present the results in Figure \ref{Figure6}.

We were surprised to find that, using FET, there is almost no difference between the number of significant comparisons made using real GO branches and using the randomly constructed ``branches''.  However, this phenomena can be understood as follows.  When calculating the significance between two gene sets, FET assumes all genes in those sets have an equal probability of being chosen.  This is a false assumption as some genes are actually more likely to be annotated to any given term in GO.  Since our random ``branches'' are constructed from actual GO annotations, these same ``high-degree'' genes are also more likely to appear in annotations made to the collection of terms defining each random ``branch.''  As noted above, experimental gene signatures also include an abundance of genes with higher levels of annotations.  Combined together, this bias means that these genes are likely to be enriched in random sets of functional categories, just because their members have more annotations overall.  We believe this illustrates a flaw of FET in that it will predict significant functional associations, not because of biological signal, but as a result of a bias in signature annotation properties.

Compared to FET, AEA finds overall fewer enriched pairs at each threshold, but, unlike FET, finds no signatures enriched in the random ``branches'', demonstrating its ability to correct for annotation biases introduced from the hierarchical relationships between those terms in the ontology.  These results give us confidence that AEA is highlighting the connections between gene sets and branches that are most likely to be truly biologically relevant and is robust against biases introduced by annotation properties.

\subsection{A Quantitative Approximation to Annotation Enrichment Analysis Partially Corrects for Annotation Bias}
\label{SPAEA}

One significant strength of AEA is that it makes no assumptions regarding the structure of gene-term annotations; however, because it uses a randomization scheme to estimate the null hypothesis, the precision of the estimated p-values is dependent upon the number of randomizations, and each run of the algorithm will give slightly different results.  Therefore, we sought an analytic approximation of AEA in order to overcome these limitations.

Given that we want to estimate the significance of annotation overlap, one logical approach is to simply count the number of annotations made to a gene set, the number of annotations made to a branch in GO, and the the number of annotations extending between that gene set and branch, and use the hypergeometric probability to determine the significance of this overlap.  We point out that this approach makes the false assumption that annotations are independent, implying that a gene could be annotated to the same term multiple times.  Because of this false assumption, it will not make as good of predictions as the randomization protocol specified by AEA; however, it can be computed quickly and without the need for any randomization to generate a null hypothesis.

Acknowledging that we are making some false assumptions regarding the structure of gene-term annotations, we propose an analytic equation, Annotation Enrichment Analysis Approximation (AEA-A).  This approximation makes use of the hypergeometric probability to calculate the significance (or p-value approximating AEA, $p_a(M_gt)$) of overlap between annotations made to a given gene set and branch in the GO hierarchy.  Given $M_g$ annotations to a gene set, $M_t$ annotations to terms belonging to a GO branch, and $M_{tot}$ annotations made in the GO ontology, the probability of finding $M_{gt}$ or more annotations in common between these two sets can be written as:
\begin{equation}
\begin{array}{lcl}
p_a(M_{gt}) & = & P(M \geq M_{gt} | M_g, M_t, M_{tot}) \vspace{2 mm} \\
& = & \displaystyle \sum_{i=M_{gt}}^{min[M_g, M_t]} \frac{{M_t \choose i} {{M_{tot}-M_t} \choose {M_g-i}}}{{M_{tot} \choose M_g}}.
\end{array}
\label{HygeoMod}
\end{equation}

We tested the performance of this approximation by determining the functional enrichment of GO terms in our randomly generated gene sets.  The results of AEA-A are uniform across varying gene set degree (see Supplemental Figure \ref{SFig2}), demonstrating that AEA-A works well at eliminating annotation bias.  However, the predicted p-values are often mis-leadingly low due to the independence assumption.  This limitation is evident in analysis performed on the experimental signatures (Figure \ref{Figure6}) -- many more comparisons are deemed ``significant'' at each threshold using AEA-A than either AEA or FET.  Furthermore, compared to AEA, the approximation is only partially able to discern between real GO branches and random ``branches''.  Therefore, we conclude that although this analytic approximation is conceptually appealing and has some advantages over FET, it does not give results that are as discerning as AEA.

\section{Discussion}

We have demonstrated that using traditional set-overlap statistics, such as FET, to evaluate the functional enrichment of gene sets is susceptible to producing false positives due to the annotation features of the GO database.  We offer a solution, Annotation Enrichment Analysis, or AEA, that fully considers these properties, eliminating any potential annotation bias in the predicted enrichment scores.  The importance of using this approach is highlighted by the fact that many published gene-signatures include a large number of highly-annotated genes.  This is likely in part due to a non-independence between identified signatures and functional annotations, since genes that are involved in a well-studied phenomena such as cancer are also more likely to be frequently annotated in these databases.  Although it is possible that newly-derived gene signatures may not exhibit the same level of annotation-bias as these previously-published signatures, it is also very probable that highly annotated genes are important in a wide variety of well-studied systems and will continue to show up and influence the results of functional enrichment analysis on newly generated gene sets.  In light of this we suggest using our approach alongside or in place of other traditional measures, especially for gene signatures that are known to contain significantly more or less annotations than one would expect by chance.  We believe that AEA will allow biologists to better interpret the functional roles of genes identified as important in their experimental system.

\section{Methods}
\label{Methods}

\subsection{Calculating Functional Enrichment using Set-Overlap Statistic}
\label{CalcFETandFDR}
In this analysis we use Fisher's Exact Test (FET) to evaluate biases in the performance of ``traditional'' functional enrichment analysis.  FET is related to the hypergeometric probability and can be used to calculate the significance, or p-value estimated using FET ($p_F(N_{gt})$) of an overlap between two independent sets.  For example, given a gene set containing $N_{g}$ genes, a GO term with $k_t$ annotations, and $N_{tot}$ total genes annotated in GO, the probability that $N_{gt}$ or more genes belong both to this gene set and are annotated to the GO term can be calculated as:
\begin{equation}
\begin{array}{lcl}
p_F(N_{gt}) & = & P(N \geq N_{gt} | N_g, k_t, N_{tot}) \vspace{2 mm} \\
& =& \displaystyle \sum_{i=N_{gt}}^{min[N_{g}, k_t]} \frac{{k_t \choose i} {{N_{tot}-k_t} \choose {N_g-i}} }{{N_{tot} \choose N_g}}.
\end{array}
\label{Hygeo}
\end{equation}
Together with the FET, we also sometimes report the False Discovery Rate (FDR) in order to account for any potential Type I error.  The FDR can be calculated as:
\begin{equation}
\beta=\alpha n / r
\label{FDR}
\end{equation}
where, $\alpha$ is the value at which an individual test was previously considered significant, $\beta$ is the value at which an individual test should be considered significant, given $n$ repetitions of that test, and $r$ is the rank of the test.

\subsection{Constructing Biased Random Gene Sets and Random ``Branches''}
\label{RandConstruction}
In order to investigate potential bias due to the annotation properties of gene sets, we constructed random gene sets with the same number of members, but with varying amounts of annotations made by those members.  Each set with a desired total number of annotations, $M_g$, was created by first randomly selecting $N_g$ genes.  We then randomly selected one gene in this gene set (gene $i$) and one gene not in the gene set (gene $j$).  If replacing gene $i$ with gene $j$ caused the total number of annotations made by genes in the gene set to approach $M_g$, we replaced gene $i$ with gene $j$ with a high probability ($p=0.95$), but if the replacement caused the average degree of the gene set to move farther away from $M_g$ we replaced gene $i$ with gene $j$ with a low probability ($p=0.05$).  This swapping continued until the total number of annotations made by the gene set was within 0.1\% of $M_g$.  In this way we created $200$ gene sets with $N_g=200$ genes each, but whose average degree ($k_{avg}=M_g/N_g$) varies from approximately $21$ to $65$.

We also constructed sets of random GO terms that represent random ``branches''.  Specifically, to mimic a branch in the GO DAG that has a parent term with $k_t$ gene annotations, we randomly ordered all the terms in GO and selected the top $N_t$ terms until the number of unique genes annotated to those $N_t$ random terms ($k_t'$) is within a small percentage of $k_t$ ($|k_t-k_t'|/k_t<0.01$).  In the case where selecting both $N_t$ and $N_{t+1}$ terms were within this limit we chose $N_t$ to minimize the absolute difference between $k_t$ and $k_t'$.  If selecting the top $N_t$ terms did not lead to a situation within this limit, we reshuffled the terms and selected the top $N_t$ terms in this new list, repeating until a suitable random collection of terms could be chosen.  In this way we created $10192$ random ``branches'' with approximately the same \emph{number} of unique genes annotated to each as to real GO branches, but in which the genes annotated to the faux parent term are influenced by a cumulation of annotations made to a random set of progeny terms.

\section*{Acknowledgements}
We would like to thank Emanuele Mazzola for helpful discussions regarding this work.

\newpage

\section*{Supplemental Figures}

\setcounter{figure}{0} \renewcommand{\thefigure}{S\arabic{figure}} 

\begin{figure}[h]
\includegraphics[height=180px]{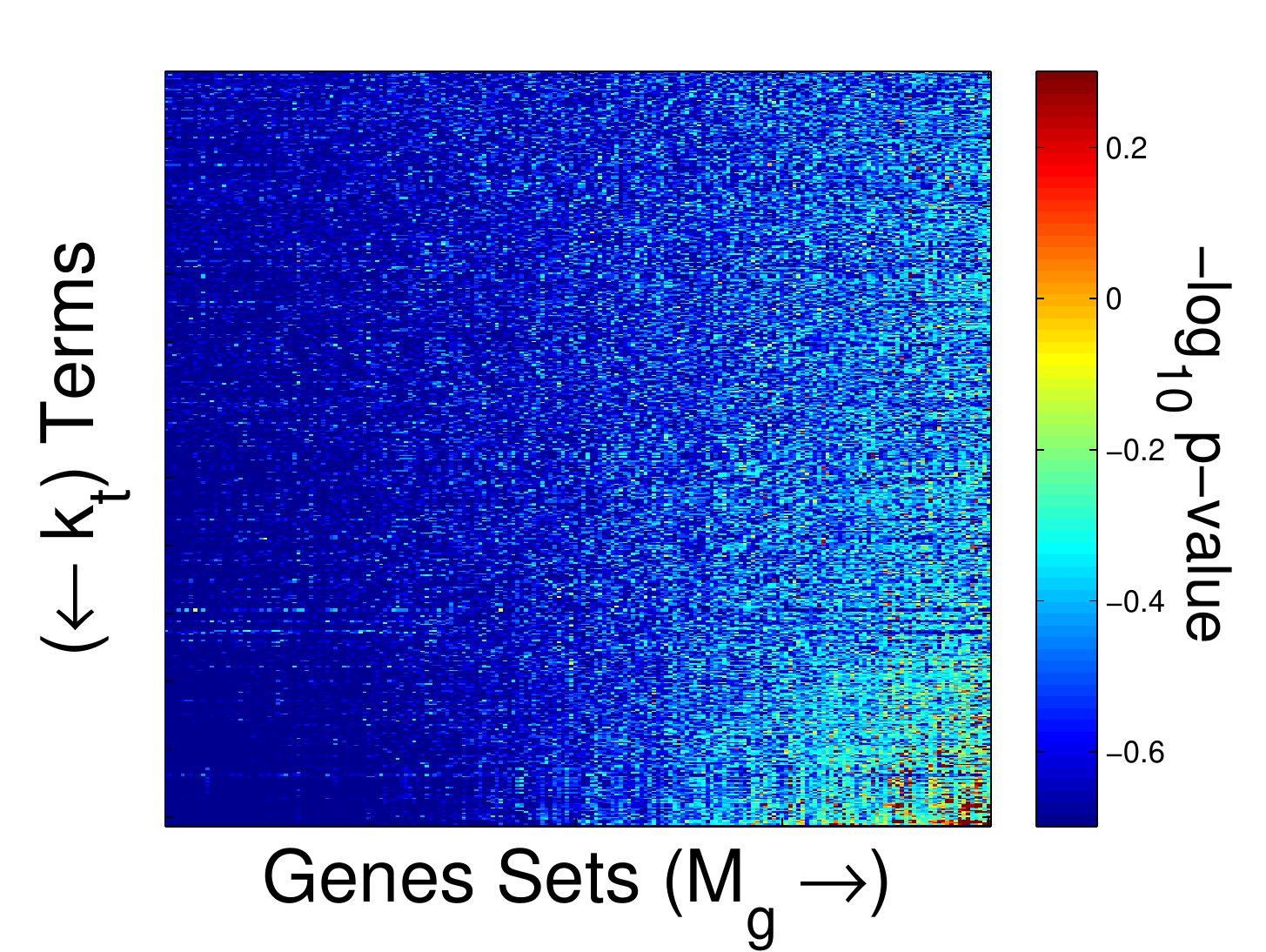}
\caption{The FDR-corrected significance of GO terms in 200 randomly generated gene sets.  The terms are ordered based on how many genes are annotated to the term ($k_t$) and the gene sets are ordered based on total the number of annotations ($M_g$) made by the $200$ genes in that set.  Although the p-values are sufficiently high that few, if any, false positives occur, there is still an obvious bias toward significant enrichment between high degree gene-set/term pairs.}
\label{SFig1}
\end{figure}

\begin{figure}[h]
\subfigure[AEA-Approximation (GO Branches)]{\label{AEAVaryK}\includegraphics[height=90px]{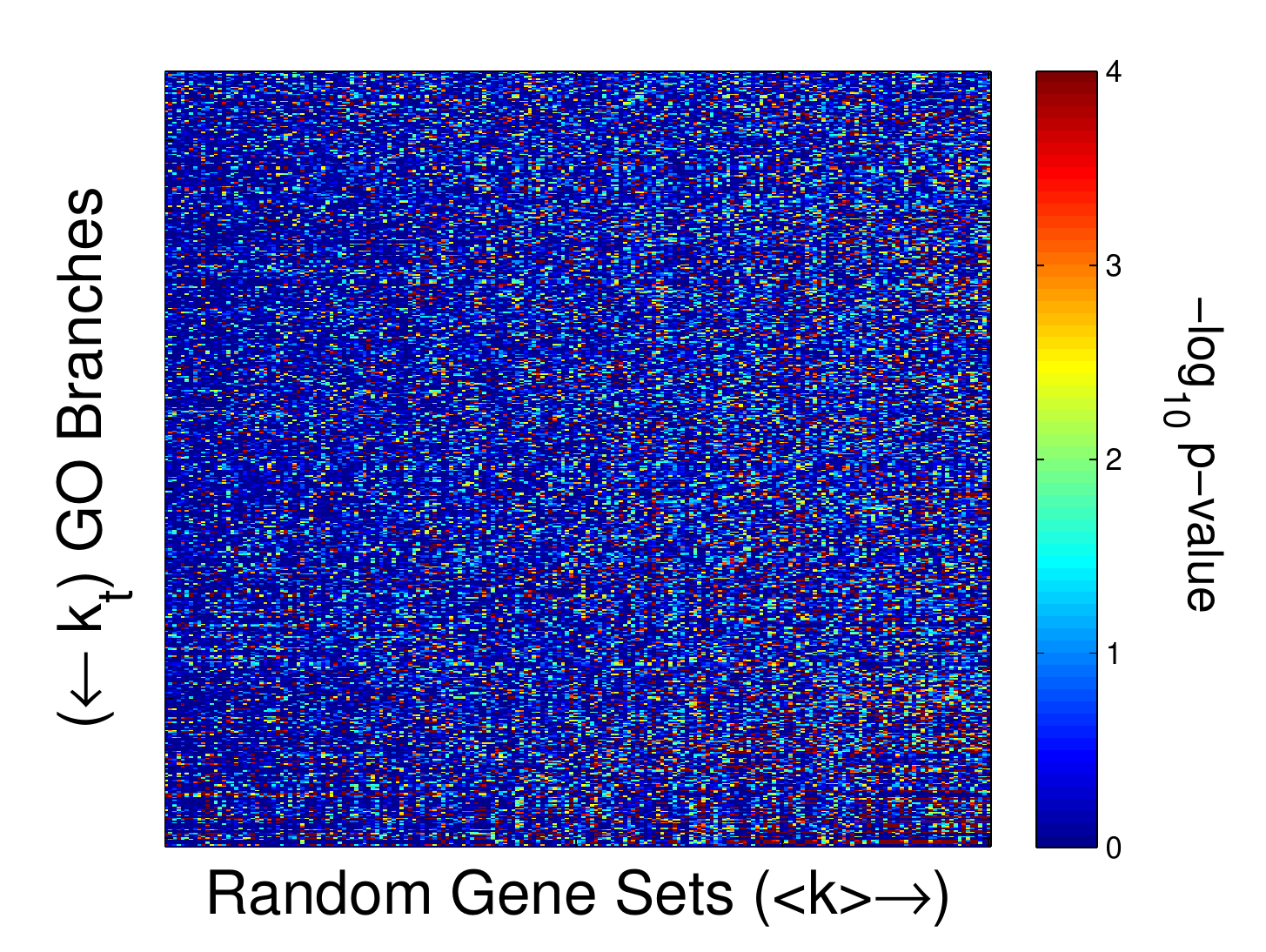}}
\subfigure[AEA-Approximation (random ``Branches'')]{\label{AEAVaryKbranch}\includegraphics[height=90px]{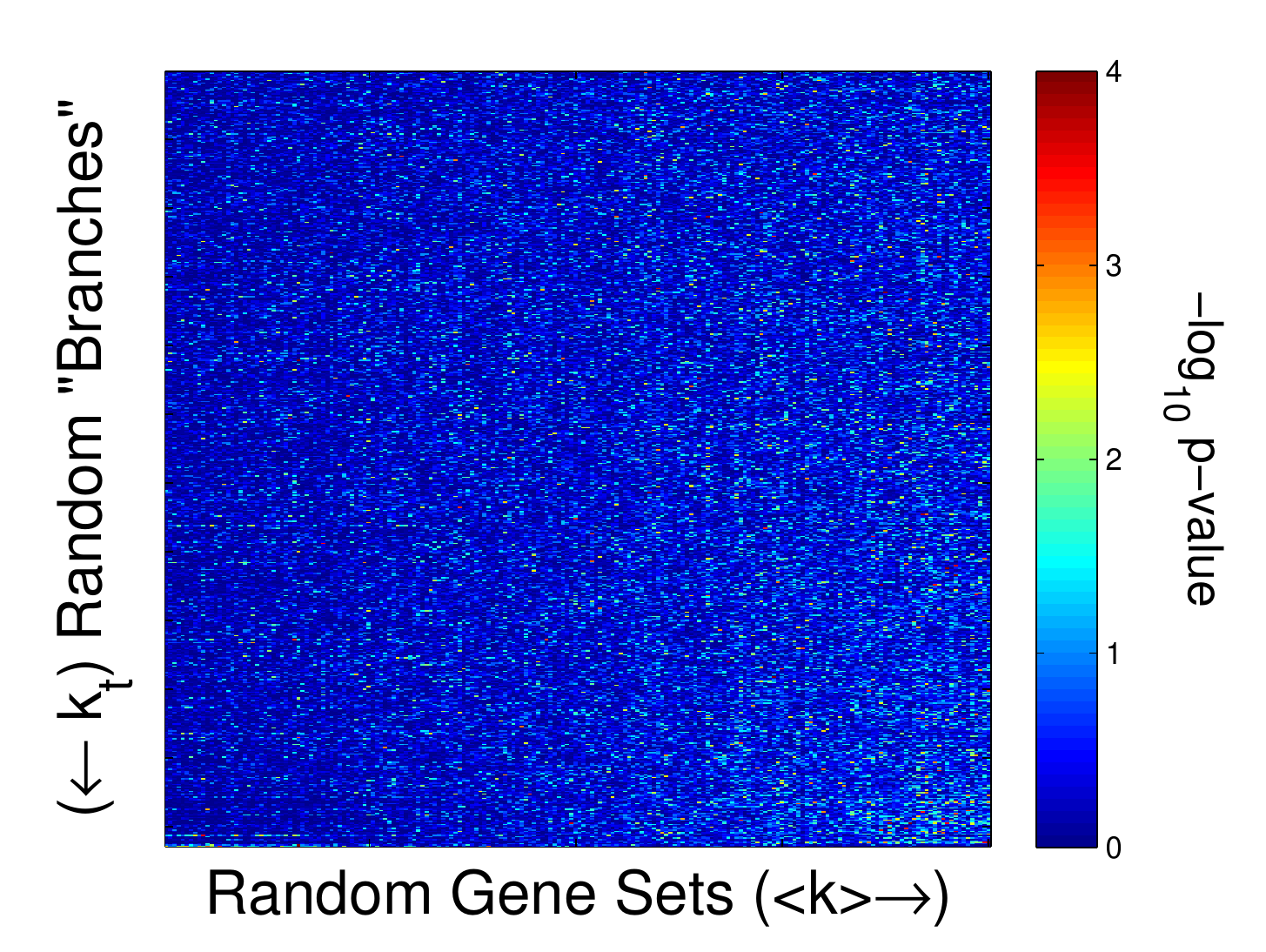}}
\caption{The significance of GO branches and random ``branches'' in 200 randomly generated gene sets according to AEA-A.  The branches are ordered based on how many genes are annotated to the parent term ($k_t$) and the gene sets are ordered based on total the number of annotations ($M_g$) made by the $200$ genes in that set.  As with AEA, AEA-A (a) successfully eliminates the annotation bias associated with high degree gene-sets and (b) predicts relatively fewer significant associations with random ``branches''.}
\label{SFig2}
\end{figure}


\begin{thebibliography}{10}
\expandafter\ifx\csname url\endcsname\relax
  \def\url#1{\texttt{#1}}\fi
\expandafter\ifx\csname urlprefix\endcsname\relax\def\urlprefix{URL }\fi
\providecommand{\bibinfo}[2]{#2}
\providecommand{\eprint}[2][]{\url{#2}}

\bibitem{DAVID}
\bibinfo{author}{Huang, D.~W.} \emph{et~al.}
\newblock \bibinfo{title}{David bioinformatics resources: expanded annotation
  database and novel algorithms to better extract biology from large gene
  lists}.
\newblock \emph{\bibinfo{journal}{Nucl. Acids Res.}}
  \textbf{\bibinfo{volume}{35}}, \bibinfo{pages}{gkm415+}
  (\bibinfo{year}{2007}).
\newblock \urlprefix\url{http://dx.doi.org/10.1093/nar/gkm415}.

\bibitem{MSigDB}
\bibinfo{author}{Subramanian, A.} \emph{et~al.}
\newblock \bibinfo{title}{Gene set enrichment analysis: A knowledge-based
  approach for interpreting genome-wide expression profiles}.
\newblock \emph{\bibinfo{journal}{Proceedings of the National Academy of
  Sciences of the United States of America}} \textbf{\bibinfo{volume}{102}},
  \bibinfo{pages}{15545--15550} (\bibinfo{year}{2005}).
\newblock \urlprefix\url{http://dx.doi.org/10.1073/pnas.0506580102}.

\bibitem{Roth03}
\bibinfo{author}{King, O.~D.}, \bibinfo{author}{Foulger, R.~E.},
  \bibinfo{author}{Dwight, S.~S.}, \bibinfo{author}{White, J.~V.} \&
  \bibinfo{author}{Roth, F.~P.}
\newblock \bibinfo{title}{Predicting gene function from patterns of
  annotation.}
\newblock \emph{\bibinfo{journal}{Genome research}}
  \textbf{\bibinfo{volume}{13}}, \bibinfo{pages}{896--904}
  (\bibinfo{year}{2003}).
\newblock \urlprefix\url{http://dx.doi.org/10.1101/gr.440803}.

\bibitem{Morris10}
\bibinfo{author}{Mostafavi, S.} \& \bibinfo{author}{Morris, Q.}
\newblock \bibinfo{title}{Fast integration of heterogeneous data sources for
  predicting gene function with limited annotation.}
\newblock \emph{\bibinfo{journal}{Bioinformatics (Oxford, England)}}
  \textbf{\bibinfo{volume}{26}}, \bibinfo{pages}{1759--1765}
  (\bibinfo{year}{2010}).
\newblock \urlprefix\url{http://dx.doi.org/10.1093/bioinformatics/btq262}.

\bibitem{GO2000}
\bibinfo{author}{Ashburner, M.} \emph{et~al.}
\newblock \bibinfo{title}{Gene ontology: tool for the unification of biology.
  the gene ontology consortium.}
\newblock \emph{\bibinfo{journal}{Nature genetics}}
  \textbf{\bibinfo{volume}{25}}, \bibinfo{pages}{25--29}
  (\bibinfo{year}{2000}).
\newblock \urlprefix\url{http://dx.doi.org/10.1038/75556}.

\bibitem{GO2010}
\bibinfo{author}{Consortium, T. G.~O.}
\newblock \bibinfo{title}{{The Gene Ontology in 2010: extensions and
  refinements}}.
\newblock \emph{\bibinfo{journal}{Nucleic Acids Research}}
  \textbf{\bibinfo{volume}{38}}, \bibinfo{pages}{D331--D335}
  (\bibinfo{year}{2010}).
\newblock \urlprefix\url{http://dx.doi.org/10.1093/nar/gkp1018}.

\bibitem{Rivals07}
\bibinfo{author}{Rivals, I.}, \bibinfo{author}{Personnaz, L.},
  \bibinfo{author}{Taing, L.} \& \bibinfo{author}{Potier, M.-C.}
\newblock \bibinfo{title}{Enrichment or depletion of a {GO} category within a
  class of genes: which test?}
\newblock \emph{\bibinfo{journal}{Bioinformatics}}
  \textbf{\bibinfo{volume}{23}}, \bibinfo{pages}{401--407}
  (\bibinfo{year}{2007}).
\newblock \urlprefix\url{http://dx.doi.org/10.1093/bioinformatics/btl633}.

\bibitem{Khatri05}
\bibinfo{author}{Khatri, P.} \& \bibinfo{author}{Dr\u{a}ghici, S.}
\newblock \bibinfo{title}{Ontological analysis of gene expression data: current
  tools, limitations, and open problems}.
\newblock \emph{\bibinfo{journal}{Bioinformatics}}
  \textbf{\bibinfo{volume}{21}}, \bibinfo{pages}{3587--3595}
  (\bibinfo{year}{2005}).
\newblock \urlprefix\url{http://dx.doi.org/10.1093/bioinformatics/bti565}.

\bibitem{Young10}
\bibinfo{author}{Young, M.~D.}, \bibinfo{author}{Wakefield, M.~J.},
  \bibinfo{author}{Smyth, G.~K.} \& \bibinfo{author}{Oshlack, A.}
\newblock \bibinfo{title}{Gene ontology analysis for {RNA}-seq: accounting for
  selection bias.}
\newblock \emph{\bibinfo{journal}{Genome biology}}
  \textbf{\bibinfo{volume}{11}}, \bibinfo{pages}{R14+} (\bibinfo{year}{2010}).
\newblock \urlprefix\url{http://dx.doi.org/10.1186/gb-2010-11-2-r14}.

\bibitem{Glass12}
\bibinfo{author}{Glass, K.}, \bibinfo{author}{Ott, E.},
  \bibinfo{author}{Losert, W.} \& \bibinfo{author}{Girvan, M.}
\newblock \bibinfo{title}{Implications of functional similarity for gene
  regulatory interactions.}
\newblock \emph{\bibinfo{journal}{Journal of the Royal Society, Interface / the
  Royal Society}}  (\bibinfo{year}{2012}).
\newblock \urlprefix\url{http://dx.doi.org/10.1098/rsif.2011.0585}.

\bibitem{GeneSigDB2012}
\bibinfo{author}{Culhane, A.~C.} \emph{et~al.}
\newblock \bibinfo{title}{{GeneSigDB}: a manually curated database and resource
  for analysis of gene expression signatures}.
\newblock \emph{\bibinfo{journal}{Nucleic Acids Research}}
  \textbf{\bibinfo{volume}{40}}, \bibinfo{pages}{D1060--D1066}
  (\bibinfo{year}{2012}).
\newblock \urlprefix\url{http://dx.doi.org/10.1093/nar/gkr901}.

\bibitem{COG}
\bibinfo{author}{Tatusov, R.~L.} \emph{et~al.}
\newblock \bibinfo{title}{The {COG} database: an updated version includes
  eukaryotes.}
\newblock \emph{\bibinfo{journal}{BMC bioinformatics}}
  \textbf{\bibinfo{volume}{4}}, \bibinfo{pages}{41+} (\bibinfo{year}{2003}).
\newblock \urlprefix\url{http://dx.doi.org/10.1186/1471-2105-4-41}.

\bibitem{KEGG}
\bibinfo{author}{Kanehisa, M.} \& \bibinfo{author}{Goto, S.}
\newblock \bibinfo{title}{{KEGG}: Kyoto encyclopedia of genes and genomes}.
\newblock \emph{\bibinfo{journal}{Nucleic Acids Research}}
  \textbf{\bibinfo{volume}{28}}, \bibinfo{pages}{27--30}
  (\bibinfo{year}{2000}).
\newblock \urlprefix\url{http://dx.doi.org/10.1093/nar/28.1.27}.

\bibitem{GenProtEC}
\bibinfo{author}{Serres, M.~H.}, \bibinfo{author}{Goswami, S.} \&
  \bibinfo{author}{Riley, M.}
\newblock \bibinfo{title}{Genprotec: an updated and improved analysis of
  functions of escherichia coli k-12 proteins.}
\newblock \emph{\bibinfo{journal}{Nucleic Acids Research}}
  \textbf{\bibinfo{volume}{32}}, \bibinfo{pages}{D300--2}
  (\bibinfo{year}{2004}).
\newblock \urlprefix\url{http://view.ncbi.nlm.nih.gov/pubmed/11471834}.

\bibitem{MetaCyc}
\bibinfo{author}{Caspi, R.} \emph{et~al.}
\newblock \bibinfo{title}{The {MetaCyc} database of metabolic pathways and
  enzymes and the {BioCyc} collection of pathway/genome databases.}
\newblock \emph{\bibinfo{journal}{Nucleic acids research}}
  \textbf{\bibinfo{volume}{38}}, \bibinfo{pages}{D473--D479}
  (\bibinfo{year}{2010}).
\newblock \urlprefix\url{http://dx.doi.org/10.1093/nar/gkp875}.

\bibitem{MultiFun}
\bibinfo{author}{Serres, M.~H.} \& \bibinfo{author}{Riley, M.}
\newblock \bibinfo{title}{{MultiFun}, a multifunctional classification scheme
  for escherichia coli k-12 gene products.}
\newblock \emph{\bibinfo{journal}{Microb Comp Genomics}}
  \textbf{\bibinfo{volume}{5}}, \bibinfo{pages}{205--222}
  (\bibinfo{year}{2000}).
\newblock \urlprefix\url{http://view.ncbi.nlm.nih.gov/pubmed/11471834}.

\bibitem{GOstat}
\bibinfo{author}{Beissbarth, T.} \& \bibinfo{author}{Speed, T.~P.}
\newblock \bibinfo{title}{{GOstat}: find statistically overrepresented gene
  ontologies within a group of genes.}
\newblock \emph{\bibinfo{journal}{Bioinformatics (Oxford, England)}}
  \textbf{\bibinfo{volume}{20}}, \bibinfo{pages}{1464--1465}
  (\bibinfo{year}{2004}).
\newblock \urlprefix\url{http://dx.doi.org/10.1093/bioinformatics/bth088}.

\bibitem{GOToolBox}
\bibinfo{author}{Martin, D.} \emph{et~al.}
\newblock \bibinfo{title}{{GOToolBox}: functional analysis of gene datasets
  based on gene ontology.}
\newblock \emph{\bibinfo{journal}{Genome Biol}} \textbf{\bibinfo{volume}{5}}
  (\bibinfo{year}{2004}).
\newblock \urlprefix\url{http://dx.doi.org/10.1186/gb-2004-5-12-r101}.

\bibitem{topGO}
\bibinfo{author}{Alexa, A.}, \bibinfo{author}{Rahnenf\"{u}hrer, J.} \&
  \bibinfo{author}{Lengauer, T.}
\newblock \bibinfo{title}{Improved scoring of functional groups from gene
  expression data by decorrelating {GO} graph structure}.
\newblock \emph{\bibinfo{journal}{Bioinformatics}}
  \textbf{\bibinfo{volume}{22}}, \bibinfo{pages}{1600--1607}
  (\bibinfo{year}{2006}).
\newblock \urlprefix\url{http://dx.doi.org/10.1093/bioinformatics/btl140}.

\bibitem{GO2001}
\bibinfo{author}{The\_gene\_ontology\_consortium}.
\newblock \bibinfo{title}{Creating the gene ontology resource: design and
  implementation.}
\newblock \emph{\bibinfo{journal}{Genome Res.}} \textbf{\bibinfo{volume}{11}},
  \bibinfo{pages}{1425--1433} (\bibinfo{year}{2001}).
\newblock \urlprefix\url{http://dx.doi.org/10.1101/gr.180801}.

\bibitem{GeneSigDB2010}
\bibinfo{author}{Culhane, A.~C.} \emph{et~al.}
\newblock \bibinfo{title}{{GeneSigDB}--a curated database of gene expression
  signatures.}
\newblock \emph{\bibinfo{journal}{Nucleic acids research}}
  \textbf{\bibinfo{volume}{38}}, \bibinfo{pages}{D716--D725}
  (\bibinfo{year}{2010}).
\newblock \urlprefix\url{http://dx.doi.org/10.1093/nar/gkp1015}.

\bibitem{Loddo09}
\bibinfo{author}{Loddo, M.} \emph{et~al.}
\newblock \bibinfo{title}{Cell-cycle-phase progression analysis identifies
  unique phenotypes of major prognostic and predictive significance in breast
  cancer}.
\newblock \emph{\bibinfo{journal}{British Journal of Cancer}}
  \textbf{\bibinfo{volume}{aop}}.
\newblock \urlprefix\url{http://dx.doi.org/10.1038/sj.bjc.6604924}.

\bibitem{Reyal08}
\bibinfo{author}{Reyal, F.} \emph{et~al.}
\newblock \bibinfo{title}{A comprehensive analysis of prognostic signatures
  reveals the high predictive capacity of the proliferation, immune response
  and {RNA} splicing modules in breast cancer.}
\newblock \emph{\bibinfo{journal}{Breast cancer research : BCR}}
  \textbf{\bibinfo{volume}{10}}, \bibinfo{pages}{R93+} (\bibinfo{year}{2008}).
\newblock \urlprefix\url{http://dx.doi.org/10.1186/bcr2192}.

\bibitem{Bohgaki05}
\bibinfo{author}{Bohgaki, T.} \emph{et~al.}
\newblock \bibinfo{title}{Up regulated expression of tumour necrosis factor
  {alpha} converting enzyme in peripheral monocytes of patients with early
  systemic sclerosis.}
\newblock \emph{\bibinfo{journal}{Annals of the rheumatic diseases}}
  \textbf{\bibinfo{volume}{64}}, \bibinfo{pages}{1165--1173}
  (\bibinfo{year}{2005}).
\newblock \urlprefix\url{http://dx.doi.org/10.1136/ard.2004.030338}.

\bibitem{Manning02}
\bibinfo{author}{Manning, G.}, \bibinfo{author}{Whyte, D.~B.},
  \bibinfo{author}{Martinez, R.}, \bibinfo{author}{Hunter, T.} \&
  \bibinfo{author}{Sudarsanam, S.}
\newblock \bibinfo{title}{The protein kinase complement of the human genome}.
\newblock \emph{\bibinfo{journal}{Science}} \textbf{\bibinfo{volume}{298}},
  \bibinfo{pages}{1912--1934} (\bibinfo{year}{2002}).
\newblock \urlprefix\url{http://dx.doi.org/10.1126/science.1075762}.

\bibitem{Dong02}
\bibinfo{author}{Dong, C.}, \bibinfo{author}{Davis, R.~J.} \&
  \bibinfo{author}{Flavell, R.~A.}
\newblock \bibinfo{title}{{MAP} kinases in the immune response.}
\newblock \emph{\bibinfo{journal}{Annual review of immunology}}
  \textbf{\bibinfo{volume}{20}}, \bibinfo{pages}{55--72}
  (\bibinfo{year}{2002}).
\newblock
  \urlprefix\url{http://dx.doi.org/10.1146/annurev.immunol.20.091301.131133}.

\bibitem{Futreal04}
\bibinfo{author}{Futreal, P.~A.} \emph{et~al.}
\newblock \bibinfo{title}{A census of human cancer genes.}
\newblock \emph{\bibinfo{journal}{Nature reviews. Cancer}}
  \textbf{\bibinfo{volume}{4}}, \bibinfo{pages}{177--183}
  (\bibinfo{year}{2004}).
\newblock \urlprefix\url{http://dx.doi.org/10.1038/nrc1299}.

\bibitem{citeulike:12015164}
\bibinfo{author}{Nepveu, A.}
\newblock \bibinfo{title}{Role of the multifunctional {CDP}/{Cut/Cux}
  homeodomain transcription factor in regulating differentiation, cell growth
  and development.}
\newblock \emph{\bibinfo{journal}{Gene}} \textbf{\bibinfo{volume}{270}},
  \bibinfo{pages}{1--15} (\bibinfo{year}{2001}).
\newblock \urlprefix\url{http://view.ncbi.nlm.nih.gov/pubmed/11403998}.

\bibitem{citeulike:12015165}
\bibinfo{author}{Magli, M.~C.}
\newblock \bibinfo{title}{The role of homeobox genes in hematopoiesis.}
\newblock \emph{\bibinfo{journal}{Biotherapy (Dordrecht, Netherlands)}}
  \textbf{\bibinfo{volume}{10}}, \bibinfo{pages}{279--294}
  (\bibinfo{year}{1998}).
\newblock \urlprefix\url{http://view.ncbi.nlm.nih.gov/pubmed/9592016}.

\bibitem{Gentles09}
\bibinfo{author}{Gentles, A.~J.} \emph{et~al.}
\newblock \bibinfo{title}{A pluripotency signature predicts histologic
  transformation and influences survival in follicular lymphoma patients.}
\newblock \emph{\bibinfo{journal}{Blood}} \textbf{\bibinfo{volume}{114}},
  \bibinfo{pages}{3158--3166} (\bibinfo{year}{2009}).
\newblock \urlprefix\url{http://dx.doi.org/10.1182/blood-2009-02-202465}.

\bibitem{Cairo08}
\bibinfo{author}{Cairo, S.} \emph{et~al.}
\newblock \bibinfo{title}{Hepatic stem-like phenotype and interplay of
  wnt/beta-catenin and myc signaling in aggressive childhood liver cancer.}
\newblock \emph{\bibinfo{journal}{Cancer cell}} \textbf{\bibinfo{volume}{14}},
  \bibinfo{pages}{471--484} (\bibinfo{year}{2008}).
\newblock \urlprefix\url{http://dx.doi.org/10.1016/j.ccr.2008.11.002}.

\bibitem{DeBerardinis08}
\bibinfo{author}{DeBerardinis, R.~J.}, \bibinfo{author}{Lum, J.~J.},
  \bibinfo{author}{Hatzivassiliou, G.} \& \bibinfo{author}{Thompson, C.~B.}
\newblock \bibinfo{title}{The biology of cancer: Metabolic reprogramming fuels
  cell growth and proliferation}.
\newblock \emph{\bibinfo{journal}{Cell Metabolism}}
  \textbf{\bibinfo{volume}{7}}, \bibinfo{pages}{11--20} (\bibinfo{year}{2008}).
\newblock \urlprefix\url{http://dx.doi.org/10.1016/j.cmet.2007.10.002}.

\bibitem{VanderHeiden09}
\bibinfo{author}{Vander~Heiden, M.~G.}, \bibinfo{author}{Cantley, L.~C.} \&
  \bibinfo{author}{Thompson, C.~B.}
\newblock \bibinfo{title}{Understanding the warburg effect: The metabolic
  requirements of cell proliferation}.
\newblock \emph{\bibinfo{journal}{Science}} \textbf{\bibinfo{volume}{324}},
  \bibinfo{pages}{1029--1033} (\bibinfo{year}{2009}).
\newblock \urlprefix\url{http://www.ncbi.nlm.nih.gov/pmc/articles/PMC2849637/}.

\bibitem{Speers09}
\bibinfo{author}{Speers, C.} \emph{et~al.}
\newblock \emph{\bibinfo{journal}{Clinical Cancer Research}}
  \bibinfo{pages}{6327--6340}.

\bibitem{Balko06}
\bibinfo{author}{Balko, J.~M.} \emph{et~al.}
\newblock \bibinfo{title}{Gene expression patterns that predict sensitivity to
  epidermal growth factor receptor tyrosine kinase inhibitors in lung cancer
  cell lines and human lung tumors}.
\newblock \emph{\bibinfo{journal}{BMC Genomics}} \textbf{\bibinfo{volume}{7}},
  \bibinfo{pages}{289+} (\bibinfo{year}{2006}).
\newblock \urlprefix\url{http://dx.doi.org/10.1186/1471-2164-7-289}.

\bibitem{Hillion08}
\bibinfo{author}{Hillion, J.} \emph{et~al.}
\newblock \bibinfo{title}{The high-mobility group a1a/signal transducer and
  activator of transcription-3 axis: an achilles heel for hematopoietic
  malignancies?}
\newblock \emph{\bibinfo{journal}{Cancer research}}
  \textbf{\bibinfo{volume}{68}}, \bibinfo{pages}{10121--10127}
  (\bibinfo{year}{2008}).
\newblock \urlprefix\url{http://dx.doi.org/10.1158/0008-5472.CAN-08-2121}.

\bibitem{Kong10}
\bibinfo{author}{Megan~Kong, Y.} \emph{et~al.}
\newblock \bibinfo{title}{Toward an ontology-based framework for clinical
  research databases.}
\newblock \emph{\bibinfo{journal}{Journal of biomedical informatics}}
  (\bibinfo{year}{2010}).
\newblock \urlprefix\url{http://dx.doi.org/10.1016/j.jbi.2010.05.001}.

\bibitem{Sotiriou03}
\bibinfo{author}{Sotiriou, C.} \emph{et~al.}
\newblock \bibinfo{title}{Breast cancer classification and prognosis based on
  gene expression profiles from a population-based study.}
\newblock \emph{\bibinfo{journal}{Proceedings of the National Academy of
  Sciences of the United States of America}} \textbf{\bibinfo{volume}{100}},
  \bibinfo{pages}{10393--10398} (\bibinfo{year}{2003}).
\newblock \urlprefix\url{http://dx.doi.org/10.1073/pnas.1732912100}.

\bibitem{MartinezDelgado05}
\bibinfo{author}{Mart\'{\i}nez-Delgado, B.} \emph{et~al.}
\newblock \bibinfo{title}{Differential expression of {NF}-{kappaB} pathway
  genes among peripheral t-cell lymphomas.}
\newblock \emph{\bibinfo{journal}{Leukemia}} \textbf{\bibinfo{volume}{19}},
  \bibinfo{pages}{2254--2263} (\bibinfo{year}{2005}).
\newblock \urlprefix\url{http://dx.doi.org/10.1038/sj.leu.2403960}.

\bibitem{Zola07}
\bibinfo{author}{Zola, H.} \emph{et~al.}
\newblock \bibinfo{title}{{CD} molecules 2006--human cell differentiation
  molecules.}
\newblock \emph{\bibinfo{journal}{Journal of immunological methods}}
  \textbf{\bibinfo{volume}{319}}, \bibinfo{pages}{1--5} (\bibinfo{year}{2007}).
\newblock \urlprefix\url{http://dx.doi.org/10.1016/j.jim.2006.11.001}.

\bibitem{Bergamaschi08}
\bibinfo{author}{Bergamaschi, A.} \emph{et~al.}
\newblock \bibinfo{title}{Extracellular matrix signature identifies breast
  cancer subgroups with different clinical outcome.}
\newblock \emph{\bibinfo{journal}{The Journal of pathology}}
  \textbf{\bibinfo{volume}{214}}, \bibinfo{pages}{357--367}
  (\bibinfo{year}{2008}).
\newblock \urlprefix\url{http://dx.doi.org/10.1002/path.2278}.

\bibitem{Moreland09}
\bibinfo{author}{Moreland, R.~T.}, \bibinfo{author}{Ryan, J.~F.},
  \bibinfo{author}{Pan, C.} \& \bibinfo{author}{Baxevanis, A.~D.}
\newblock \bibinfo{title}{The homeodomain resource: a comprehensive collection
  of sequence, structure, interaction, genomic and functional information on
  the homeodomain protein family.}
\newblock \emph{\bibinfo{journal}{Database : the journal of biological
  databases and curation}} \textbf{\bibinfo{volume}{2009}}
  (\bibinfo{year}{2009}).
\newblock \urlprefix\url{http://dx.doi.org/10.1093/database/bap004}.

\bibitem{Yang08}
\bibinfo{author}{Yang, S.~X.} \emph{et~al.}
\newblock \bibinfo{title}{Gene expression profile and angiogenic marker
  correlates with response to neoadjuvant bevacizumab followed by bevacizumab
  plus chemotherapy in breast cancer.}
\newblock \emph{\bibinfo{journal}{Clinical cancer research : an official
  journal of the American Association for Cancer Research}}
  \textbf{\bibinfo{volume}{14}}, \bibinfo{pages}{5893--5899}
  (\bibinfo{year}{2008}).
\newblock \urlprefix\url{http://dx.doi.org/10.1158/1078-0432.CCR-07-4762}.

\bibitem{VandenEynden07}
\bibinfo{author}{Van~den Eynden, G.~G.} \emph{et~al.}
\newblock \bibinfo{title}{Differential expression of hypoxia and
  (lymph)angiogenesis-related genes at different metastatic sites in breast
  cancer.}
\newblock \emph{\bibinfo{journal}{Clinical \& experimental metastasis}}
  \textbf{\bibinfo{volume}{24}}, \bibinfo{pages}{13--23}
  (\bibinfo{year}{2007}).
\newblock \urlprefix\url{http://dx.doi.org/10.1007/s10585-006-9049-3}.

\bibitem{Rae04}
\bibinfo{author}{Rae, M.~T.} \emph{et~al.}
\newblock \bibinfo{title}{Steroid signalling in human ovarian surface
  epithelial cells: the response to interleukin-1alpha determined by microarray
  analysis.}
\newblock \emph{\bibinfo{journal}{The Journal of endocrinology}}
  \textbf{\bibinfo{volume}{183}}, \bibinfo{pages}{19--28}
  (\bibinfo{year}{2004}).
\newblock \urlprefix\url{http://dx.doi.org/10.1677/joe.1.05754}.

\bibitem{Wood10}
\bibinfo{author}{Wood, C.~E.}, \bibinfo{author}{Kaplan, J.~R.},
  \bibinfo{author}{Fontenot, M.~B.}, \bibinfo{author}{Williams, J.~K.} \&
  \bibinfo{author}{Cline, J.~M.}
\newblock \bibinfo{title}{Endometrial profile of tamoxifen and low-dose
  estradiol combination therapy.}
\newblock \emph{\bibinfo{journal}{Clinical cancer research : an official
  journal of the American Association for Cancer Research}}
  \textbf{\bibinfo{volume}{16}}, \bibinfo{pages}{946--956}
  (\bibinfo{year}{2010}).
\newblock \urlprefix\url{http://dx.doi.org/10.1158/1078-0432.CCR-09-1541}.

\bibitem{Hassan09}
\bibinfo{author}{Hassan, K.~A.}, \bibinfo{author}{Chen, G.},
  \bibinfo{author}{Kalemkerian, G.~P.}, \bibinfo{author}{Wicha, M.~S.} \&
  \bibinfo{author}{Beer, D.~G.}
\newblock \bibinfo{title}{An embryonic stem cell-like signature identifies
  poorly differentiated lung adenocarcinoma but not squamous cell carcinoma}.
\newblock \emph{\bibinfo{journal}{Clinical Cancer Research}}
  \textbf{\bibinfo{volume}{15}}, \bibinfo{pages}{6386--6390}
  (\bibinfo{year}{2009}).
\newblock \urlprefix\url{http://dx.doi.org/10.1158/1078-0432.CCR-09-1105}.

\bibitem{Mason04}
\bibinfo{author}{Mason, D.~X.}, \bibinfo{author}{Jackson, T.~J.} \&
  \bibinfo{author}{Lin, A.~W.}
\newblock \bibinfo{title}{Molecular signature of oncogenic ras-induced
  senescence.}
\newblock \emph{\bibinfo{journal}{Oncogene}} \textbf{\bibinfo{volume}{23}},
  \bibinfo{pages}{9238--9246} (\bibinfo{year}{2004}).
\newblock \urlprefix\url{http://dx.doi.org/10.1038/sj.onc.1208172}.

\bibitem{Lauss08}
\bibinfo{author}{Lauss, M.} \emph{et~al.}
\newblock \bibinfo{title}{Consensus genes of the literature to predict breast
  cancer recurrence.}
\newblock \emph{\bibinfo{journal}{Breast cancer research and treatment}}
  \textbf{\bibinfo{volume}{110}}, \bibinfo{pages}{235--244}
  (\bibinfo{year}{2008}).
\newblock \urlprefix\url{http://dx.doi.org/10.1007/s10549-007-9716-3}.

\bibitem{Mutarelli08}
\bibinfo{author}{Mutarelli, M.} \emph{et~al.}
\newblock \bibinfo{title}{Time-course analysis of genome-wide gene expression
  data from hormone-responsive human breast cancer cells.}
\newblock \emph{\bibinfo{journal}{BMC bioinformatics}}
  \textbf{\bibinfo{volume}{9 Suppl 2}}, \bibinfo{pages}{S12+}
  (\bibinfo{year}{2008}).
\newblock \urlprefix\url{http://dx.doi.org/10.1186/1471-2105-9-S2-S12}.

\bibitem{Williams09}
\bibinfo{author}{Williams, C.~M.} \emph{et~al.}
\newblock \bibinfo{title}{{AP}-2gamma promotes proliferation in breast tumour
  cells by direct repression of the {CDKN1A} gene.}
\newblock \emph{\bibinfo{journal}{The EMBO journal}}
  \textbf{\bibinfo{volume}{28}}, \bibinfo{pages}{3591--3601}
  (\bibinfo{year}{2009}).
\newblock \urlprefix\url{http://dx.doi.org/10.1038/emboj.2009.290}.

\bibitem{Sotiriou06}
\bibinfo{author}{Sotiriou, C.} \emph{et~al.}
\newblock \bibinfo{title}{Gene expression profiling in breast cancer:
  understanding the molecular basis of histologic grade to improve prognosis.}
\newblock \emph{\bibinfo{journal}{Journal of the National Cancer Institute}}
  \textbf{\bibinfo{volume}{98}}, \bibinfo{pages}{262--272}
  (\bibinfo{year}{2006}).
\newblock \urlprefix\url{http://dx.doi.org/10.1093/jnci/djj052}.

\bibitem{Thorner09}
\bibinfo{author}{Thorner, A.~R.} \emph{et~al.}
\newblock \bibinfo{title}{In vitro and in vivo analysis of {B-Myb} in
  basal-like breast cancer.}
\newblock \emph{\bibinfo{journal}{Oncogene}} \textbf{\bibinfo{volume}{28}},
  \bibinfo{pages}{742--751} (\bibinfo{year}{2009}).
\newblock \urlprefix\url{http://dx.doi.org/10.1038/onc.2008.430}.

\bibitem{Loi08}
\bibinfo{author}{Loi, S.} \emph{et~al.}
\newblock \bibinfo{title}{Predicting prognosis using molecular profiling in
  estrogen receptor-positive breast cancer treated with tamoxifen.}
\newblock \emph{\bibinfo{journal}{BMC genomics}} \textbf{\bibinfo{volume}{9}},
  \bibinfo{pages}{239+} (\bibinfo{year}{2008}).
\newblock \urlprefix\url{http://dx.doi.org/10.1186/1471-2164-9-239}.

\bibitem{Miller05}
\bibinfo{author}{Miller, L.~D.} \emph{et~al.}
\newblock \bibinfo{title}{An expression signature for p53 status in human
  breast cancer predicts mutation status, transcriptional effects, and patient
  survival}.
\newblock \emph{\bibinfo{journal}{Proceedings of the National Academy of
  Sciences of the United States of America}} \textbf{\bibinfo{volume}{102}},
  \bibinfo{pages}{13550--13555} (\bibinfo{year}{2005}).
\newblock \urlprefix\url{http://dx.doi.org/10.1073/pnas.0506230102}.

\bibitem{Wirapati08}
\bibinfo{author}{Wirapati, P.} \emph{et~al.}
\newblock \bibinfo{title}{Meta-analysis of gene expression profiles in breast
  cancer: toward a unified understanding of breast cancer subtyping and
  prognosis signatures}.
\newblock \emph{\bibinfo{journal}{Breast Cancer Research}}
  \textbf{\bibinfo{volume}{10}}, \bibinfo{pages}{R65+} (\bibinfo{year}{2008}).
\newblock \urlprefix\url{http://dx.doi.org/10.1186/bcr2124}.

\bibitem{Crawford08}
\bibinfo{author}{Crawford, N.~P.} \emph{et~al.}
\newblock \bibinfo{title}{Bromodomain 4 activation predicts breast cancer
  survival.}
\newblock \emph{\bibinfo{journal}{Proceedings of the National Academy of
  Sciences of the United States of America}} \textbf{\bibinfo{volume}{105}},
  \bibinfo{pages}{6380--6385} (\bibinfo{year}{2008}).
\newblock \urlprefix\url{http://dx.doi.org/10.1073/pnas.0710331105}.

\bibitem{Troester06}
\bibinfo{author}{Troester, M.~A.} \emph{et~al.}
\newblock \bibinfo{title}{Gene expression patterns associated with p53 status
  in breast cancer.}
\newblock \emph{\bibinfo{journal}{BMC cancer}} \textbf{\bibinfo{volume}{6}},
  \bibinfo{pages}{276+} (\bibinfo{year}{2006}).
\newblock \urlprefix\url{http://dx.doi.org/10.1186/1471-2407-6-276}.

\bibitem{Dairkee04}
\bibinfo{author}{Dairkee, S.~H.} \emph{et~al.}
\newblock \bibinfo{title}{A molecular 'signature' of primary breast cancer
  cultures; patterns resembling tumor tissue.}
\newblock \emph{\bibinfo{journal}{BMC genomics}} \textbf{\bibinfo{volume}{5}}
  (\bibinfo{year}{2004}).
\newblock \urlprefix\url{http://dx.doi.org/10.1186/1471-2164-5-47}.

\bibitem{Chang04}
\bibinfo{author}{Chang, H.~Y.} \emph{et~al.}
\newblock \bibinfo{title}{Gene expression signature of fibroblast serum
  response predicts human cancer progression: similarities between tumors and
  wounds.}
\newblock \emph{\bibinfo{journal}{PLoS biology}} \textbf{\bibinfo{volume}{2}},
  \bibinfo{pages}{e7+} (\bibinfo{year}{2004}).
\newblock \urlprefix\url{http://dx.doi.org/10.1371/journal.pbio.0020007}.

\bibitem{VonKopylow10}
\bibinfo{author}{von Kopylow, K.} \emph{et~al.}
\newblock \bibinfo{title}{Screening for biomarkers of spermatogonia within the
  human testis: a whole genome approach.}
\newblock \emph{\bibinfo{journal}{Human reproduction (Oxford, England)}}
  \textbf{\bibinfo{volume}{25}}, \bibinfo{pages}{1104--1112}
  (\bibinfo{year}{2010}).
\newblock \urlprefix\url{http://dx.doi.org/10.1093/humrep/deq053}.

\bibitem{Tome05}
\bibinfo{author}{Tome, M.~E.} \emph{et~al.}
\newblock \bibinfo{title}{A redox signature score identifies diffuse large
  b-cell lymphoma patients with a poor prognosis.}
\newblock \emph{\bibinfo{journal}{Blood}} \textbf{\bibinfo{volume}{106}},
  \bibinfo{pages}{3594--3601} (\bibinfo{year}{2005}).
\newblock \urlprefix\url{http://dx.doi.org/10.1182/blood-2005-02-0487}.

\bibitem{Piccaluga07}
\bibinfo{author}{Piccaluga, P. P.~P.} \emph{et~al.}
\newblock \bibinfo{title}{Gene expression analysis of peripheral t cell
  lymphoma, unspecified, reveals distinct profiles and new potential
  therapeutic targets.}
\newblock \emph{\bibinfo{journal}{The Journal of clinical investigation}}
  \textbf{\bibinfo{volume}{117}}, \bibinfo{pages}{823--834}
  (\bibinfo{year}{2007}).
\newblock \urlprefix\url{http://dx.doi.org/10.1172/JCI26833}.

\bibitem{Chin09}
\bibinfo{author}{Chin, M.}, \bibinfo{author}{Herscovitch, M.},
  \bibinfo{author}{Zhang, N.}, \bibinfo{author}{Waxman, D.~J.} \&
  \bibinfo{author}{Gilmore, T.~D.}
\newblock \bibinfo{title}{Overexpression of an activated {REL} mutant enhances
  the transformed state of the human b-lymphoma {BJAB} cell line and alters its
  gene expression profile.}
\newblock \emph{\bibinfo{journal}{Oncogene}} \textbf{\bibinfo{volume}{28}},
  \bibinfo{pages}{2100--2111} (\bibinfo{year}{2009}).
\newblock \urlprefix\url{http://dx.doi.org/10.1038/onc.2009.74}.

\bibitem{Leval07}
\bibinfo{author}{de~Leval, L.} \emph{et~al.}
\newblock \bibinfo{title}{The gene expression profile of nodal peripheral
  t-cell lymphoma demonstrates a molecular link between angioimmunoblastic
  t-cell lymphoma ({AITL}) and follicular helper t ({TFH}) cells.}
\newblock \emph{\bibinfo{journal}{Blood}} \textbf{\bibinfo{volume}{109}},
  \bibinfo{pages}{4952--4963} (\bibinfo{year}{2007}).
\newblock \urlprefix\url{http://dx.doi.org/10.1182/blood-2006-10-055145}.

\bibitem{Bertucci02}
\bibinfo{author}{Bertucci, F.} \emph{et~al.}
\newblock \bibinfo{title}{Prognosis of breast cancer and gene expression
  profiling using {DNA} arrays.}
\newblock \emph{\bibinfo{journal}{Annals of the New York Academy of Sciences}}
  \textbf{\bibinfo{volume}{975}}, \bibinfo{pages}{217--231}
  (\bibinfo{year}{2002}).
\newblock \urlprefix\url{http://view.ncbi.nlm.nih.gov/pubmed/12538167}.

\bibitem{Scott04}
\bibinfo{author}{Scott, L.~A.} \emph{et~al.}
\newblock \bibinfo{title}{Invasion of normal human fibroblasts induced by
  {v-Fos} is independent of proliferation, immortalization, and the tumor
  suppressors {p16INK4a} and p53.}
\newblock \emph{\bibinfo{journal}{Molecular and cellular biology}}
  \textbf{\bibinfo{volume}{24}}, \bibinfo{pages}{1540--1559}
  (\bibinfo{year}{2004}).
\newblock \urlprefix\url{http://www.ncbi.nlm.nih.gov/pmc/articles/PMC344183/}.

\bibitem{Godard03}
\bibinfo{author}{Godard, S.} \emph{et~al.}
\newblock \bibinfo{title}{Classification of human astrocytic gliomas on the
  basis of gene expression: a correlated group of genes with angiogenic
  activity emerges as a strong predictor of subtypes.}
\newblock \emph{\bibinfo{journal}{Cancer research}}
  \textbf{\bibinfo{volume}{63}}, \bibinfo{pages}{6613--6625}
  (\bibinfo{year}{2003}).
\newblock \urlprefix\url{http://view.ncbi.nlm.nih.gov/pubmed/14583454}.

\bibitem{Kocer08}
\bibinfo{author}{Ko\c{c}er, S.~S.}, \bibinfo{author}{Djuri\'{c}, P.~M.},
  \bibinfo{author}{Bugallo, M.~F.}, \bibinfo{author}{Simon, S.~R.} \&
  \bibinfo{author}{Matic, M.}
\newblock \bibinfo{title}{Transcriptional profiling of putative human
  epithelial stem cells.}
\newblock \emph{\bibinfo{journal}{BMC genomics}} \textbf{\bibinfo{volume}{9}},
  \bibinfo{pages}{359+} (\bibinfo{year}{2008}).
\newblock \urlprefix\url{http://dx.doi.org/10.1186/1471-2164-9-359}.

\bibitem{Chandran07}
\bibinfo{author}{Chandran, U.~R.} \emph{et~al.}
\newblock \bibinfo{title}{Gene expression profiles of prostate cancer reveal
  involvement of multiple molecular pathways in the metastatic process.}
\newblock \emph{\bibinfo{journal}{BMC cancer}} \textbf{\bibinfo{volume}{7}},
  \bibinfo{pages}{64+} (\bibinfo{year}{2007}).
\newblock \urlprefix\url{http://dx.doi.org/10.1186/1471-2407-7-64}.

\bibitem{Jenner03}
\bibinfo{author}{Jenner, R.~G.} \emph{et~al.}
\newblock \bibinfo{title}{Kaposi's sarcoma-associated herpesvirus-infected
  primary effusion lymphoma has a plasma cell gene expression profile.}
\newblock \emph{\bibinfo{journal}{Proceedings of the National Academy of
  Sciences of the United States of America}} \textbf{\bibinfo{volume}{100}},
  \bibinfo{pages}{10399--10404} (\bibinfo{year}{2003}).
\newblock \urlprefix\url{http://dx.doi.org/10.1073/pnas.1630810100}.

\end{thebibliography}
\end{document}